\documentclass[11pt,a4paper,english]{article}

\usepackage{graphicx}
\usepackage{amssymb,latexsym,cite}
\usepackage{amsmath}
\usepackage{amsfonts}
\usepackage{mathrsfs}
\usepackage{bbm}
\usepackage{bm}
\usepackage[T1]{fontenc}

\usepackage{hhline}

\usepackage{epsf}
\usepackage{pdfsync}
\usepackage{epsfig}
\usepackage[parfill]{parskip}
\usepackage[matrix,arrow,color]{xy}
\usepackage[usenames]{color}
\usepackage{hyperref}
\usepackage{bbm}

\input{xy}
\xyoption{all}

\input{epsf}

\makeatletter

\catcode`@=11

\numberwithin{equation}{section}
\catcode`@=12

\newcommand{\ba}{\begin{eqnarray*}}
\newcommand{\ea}{\end{eqnarray*}}
\newcommand{\ban}{\begin{eqnarray}}
\newcommand{\ean}{\end{eqnarray}}

\newcommand{\IZ}{\mathbb{Z}}
\newcommand{\IC}{\mathbb{C}}

\newcommand{\IR}{\mathbb{R}}

\newcommand{\wJ}{\widetilde{J}{}}
\newcommand{\weta}{\widetilde{\zeta}{}}
\newcommand{\wx}{\widetilde{x}{}}
\newcommand{\wy}{\widetilde{y}{}}
\newcommand{\wG}{\widetilde{G}{}}
\newcommand{\wM}{\widetilde{M}{}}
\newcommand{\wcM}{\widetilde{\cM}{}}

\newcommand{\bX}{\mathbb{X}}

\newcommand{\cW}{{\mathcal W}}

\newcommand{\cM}{{\mathcal M}}

\newcommand{\cH}{{\mathcal H}}
\newcommand{\cA}{{\mathcal A}}

\newcommand{\cO}{{\mathcal O}}

\newcommand{\cK}{{\mathcal K}}
\newcommand{\cP}{{\mathcal P}}

\newcommand{\cJ}{{\mathcal J}}

\newcommand{\ccX}{{\mathscr X}}
\newcommand{\ccG}{{\mathscr G}}

\newcommand{\sfM}{{\mathsf{M}}}

\newcommand{\unit}{\mathbbm{1}}   			

\newcommand{\mbf}[1]{{\boldsymbol {#1} }}

\def\e{{\,\rm e}\,}

\def\ii{{\,{\rm i}\,}}
\def\dd{{\rm d}}

\def\beq{\begin{equation}}
\def\bee{\begin{equation}}
\def\eeq{\end{equation}}
\def\bea{\begin{eqnarray}}
\def\eea{\end{eqnarray}}
\def\bd{\begin{displaymath}}
\def\ed{\end{displaymath}}

\newcommand{\Cint}{\int\kern-10.5pt-\kern7pt}

\newcommand{\be}{\begin{equation}}
\newcommand{\ee}{\end{equation}}

\newcommand\fverbit{\egroup\item[\fbox{\unhbox\pippobox}]}
\newbox\pippobox

{

\def\be{\begin{equation}}
\def\ee{\end{equation}}
\def\bea{\begin{eqnarray}}
\def\eea{\end{eqnarray}}

\begin{document}

\begin{titlepage}
\setcounter{page}{0}
\begin{flushright}
\small
{\sf Imperial--TP--2019--CH--02}\\
{\sf EMPG--19--04}
\normalsize
\end{flushright}

\vskip 1.8cm

\begin{center}

{\Large\bf Noncommutative gauge theories \\[2mm] on D-branes in non-geometric backgrounds}

\vspace{15mm}

{\large\bf Chris~Hull$^{(a)}$} \ and \ 
{\large\bf Richard~J.~Szabo$^{(b)}$}
\\[6mm]

\noindent{\em $^{(a)}$ The
Blackett Laboratory\\ Imperial College London\\ Prince Consort Road, London SW7 2AZ, UK} \\ Email: \ {\tt
  c.hull@imperial.ac.uk}\\[4mm]
\noindent{\em $^{(b)}$ Department of Mathematics\\ Heriot-Watt
  University\\
Colin Maclaurin Building, Riccarton, Edinburgh EH14 4AS, UK\\ 
Maxwell Institute for Mathematical Sciences, Edinburgh, UK\\
The Higgs Centre for Theoretical Physics, Edinburgh, UK}\\
Email: \ {\tt
  R.J.Szabo@hw.ac.uk}

\vspace{20mm}

\begin{abstract}
\noindent
We investigate the noncommutative  gauge theories arising on the worldvolumes of
D-branes  in non-geometric backgrounds obtained by T-duality from
twisted tori. We revisit the
low-energy effective description of D-branes on three-dimensional T-folds, examining
both cases of parabolic and elliptic twists in detail.
We give a detailed description of the decoupling limits and  explore various
physical consequences of the open string non-geometry. The T-duality
monodromies of the non-geometric backgrounds lead to Morita duality
monodromies of the noncommutative Yang--Mills theories induced on the
D-branes. While the parabolic twists recover the well-known examples
of noncommutative principal torus bundles from topological T-duality, the elliptic twists give
new examples of noncommutative fibrations with non-geometric torus fibres.
We extend these considerations
to D-branes in backgrounds with $R$-flux, using the doubled geometry formulation, finding that both the  non-geometric background and the D-brane gauge theory necessarily have explicit dependence on the dual coordinates, and so have no conventional formulation in spacetime.
\end{abstract}

\end{center}


\end{titlepage}

\newpage


\tableofcontents

\bigskip

\section{Introduction and summary} \label{intro}

One of the most striking features of T-duality, which relates
different string backgrounds describing the same physics, is that it leads to the
possibility of non-geometric backgrounds which do not
have a description in terms of conventional Riemannian
geometry \cite{Hull2004} (see
e.g.~\cite{Wecht2007,Berman2013,Plauschinn2018} for reviews and
further references). 
Some non-geometric backgrounds arise as T-duals of conventional geometric backgrounds, while others are not   geometric in any duality frame.
Typical examples start with geometric spaces which admit a torus
fibration, with transition functions that are diffeomorphisms of the
torus fibres and shifts of the $B$-field. T-duality transformations along the torus fibres (using
the standard Buscher rules~\cite{Buscher1987,Buscher1988} on a covering space)
in general lead to T-folds \cite{Hull2004}. These are locally geometric -- locally they look like a product of the torus with a patch of the base -- but the transition functions in general involve T-duality transformations on the torus fibres. 
The Buscher rules give T-duality in isometric directions.
For non-isometric directions, there is a notion of generalised T-duality that can be applied~\cite{Dabholkar2005}.
For a circular direction in which the fields depend explicitly on the coordinate $x$ of that circle, 
a generalised T-duality transforms this to a configuration in which the fields depend on the coordinate $\wx$ of the T-dual circle. This $\wx$-dependence means that it cannot be viewed as a conventional background even locally, but has intrinsic dependence on the T-dual coordinates, so a doubled geometry formulation is essential.
This concept of generalised T-duality 
has been checked in asymmetric orbifold limits~\cite{Dabholkar2005}, and is in
agreement with the concept of generalised T-duality  arising in double field theory \cite{Hull:2009mi,Hohm:2010jy,Hohm:2010pp}.  {We will refer to  configurations 
in which fields and/or transition functions have
 explicit dependence on the dual coordinates $\wx$ as  {\it essentially doubled}.}
 
In this paper we will consider $n$-dimensional backgrounds obtained by T-dualising the 
simplest examples of
torus bundles, 
 which are fibrations of $n-1$-dimensional tori $T^{n-1}$ over
a circle $S^1$ with vanishing $B$-field, sometimes referred to as 
 twisted tori~\cite{Dabholkar2005,Dabholkar2002,Kachru2002,Hull:1998vy,
Shelton2005,Shelton2006}.\footnote{Closely related non-geometric backgrounds involving torus bundles with T-duality monodromy around singular fibres were discussed in 
\cite{Hellerman2002,Flournoy2004}.}
  The monodromy around the base circle is a diffeomorphism of the torus fibres, in the mapping class group $SL(n-1, \mathbb{Z})$. 
 These and their T-duals give compactifications with a duality
 twist~\cite{Dabholkar2002}, which are stringy generalisations of
 Scherk--Schwarz reductions~\cite{Hull:1998vy}. For definiteness, we will focus on the
 case of backgrounds in $n=3$ dimensions, where all of our
 considerations can be made explicit. Then the simplest case is that of a parabolic
monodromy, acting as an integer shift $\tau \mapsto \tau +m$ of the
complex structure modulus $\tau$ of the two-torus $T^2$. In this case,   the torus bundle is the nilfold of degree $m$, which is T-dual to a geometric three-torus $T^3$ with
$H$-flux of the $B$-field proportional to $m$~\cite{Hull:1998vy}.
Applying T-duality transformations then results in a much-studied chain
of transformations between geometric and non-geometric backgrounds~\cite{Dabholkar2002,Kachru2002,
Shelton2005,Dabholkar2005,Shelton2006}. 
This is conventionally depicted in a schematic form as~\cite{Shelton2005}
\bea\label{eq:Tdualitychain}
{H_{ijk}} \ \stackrel{{\sf T}_i}{\longleftrightarrow} \ {f^i{}_{jk}} \
\stackrel{{\sf T}_j}{\longleftrightarrow} \ {Q^{ij}{}_k} \
\stackrel{{\sf T}_k}{\longleftrightarrow}\ {R^{ijk}} 
\eea
where ${\sf T}_i$ denotes a T-duality transformation along the $i$-th
coordinate direction. 
Successive T-dualities take the three-torus with $H$-flux to a nilfold with what is sometimes called ``geometric flux''~$f$, then to a T-fold with ``$Q$-flux'', and finally a generalised T-duality takes this to  {an essentially doubled  space} with ``$R$-flux''.
{The cases with $f$-, $H$- and $Q$-flux can be thought of as $T^2$
  conformal field theories fibred over a circle coordinate $x$, with monodromy in the T-duality group $O(2,2;\mathbb{Z})$, while the case with $R$-flux is an essentially doubled  space
 which is a fibration over the T-dual circle with dual coordinate $\wx$ and monodromy in the T-duality group.}
For more general monodromies, such as
the elliptic monodromies that we consider in detail below, the
results are rather different and do not follow the pattern suggested by~\eqref{eq:Tdualitychain}. 
{As we shall see for the elliptic case, acting on the twisted torus \lq with
$f$-flux'  with either  ${\sf T}_i$ or ${\sf T}_j$ gives a T-fold, and no dual with only $H$-flux arises. A further T-duality then gives an
essentially doubled  space.}

A useful perspective for understanding non-geometry in string theory
is to study   D-branes in these backgrounds. D-branes can be used as
probes to analyse the geometry of a string background and to provide an alternative
definition of the background geometry in terms of the moduli space of
the probe. D-branes in non-geometric backgrounds were previously
discussed from the point of view of doubled (twisted) torus geometry
in~\cite{Hull2004,Lawrence2006,Albertsson2008,Albertsson2011},
directly in string theory from a target space perspective
in~\cite{Lowe2003,Ellwood2006,Grange2006}, and from open string
worldsheet theory in~\cite{Kawai2007,Cordonier2018}.

In the present paper we focus on an approach based on effective
field theory, reinterpreting all of the T-duality transformations in the
chain \eqref{eq:Tdualitychain} for the nilfold and the corresponding chains for other backgrounds
in terms of open
strings. In this setting it is important to define a low-energy
scaling limit which decouples the 
deformations of geometry due to non-locality of strings from the ``genuine'' non-geometry due to background fields. In the
case of D-branes in flat space with a constant $B$-field background, 
the D-brane
worldvolume theory is a
noncommutative supersymmetric Yang--Mills theory~\cite{Douglas1997}, and the 
  decoupling
limit was carefully set out in~\cite{Seiberg1999}; these considerations were extended to the case of
D-branes in curved backgrounds with non-constant $B$-fields and non-zero $H$-flux
in~\cite{Cornalba2001,Herbst2001} (see
e.g.~\cite{Douglas2001,Szabo2001,Szabo2006} for reviews and further references). This limit is often neglected in the literature on
embeddings of worldvolume noncommutative gauge theories in string
theory. 

To this end, we revisit the problem of formulating effective
noncommutative Yang--Mills theories on D-branes in non-geometric tori
in the decoupling limit, extending the earlier work
of~\cite{Ellwood2006,Grange2006} (see also~\cite{Lowe2003}) in various
directions. In these works, D3-branes in the simplest T-fold
background, originating via T-duality from a flat three-torus with $H$-flux, are shown to have an effective description as a
noncommutative gauge theory on a flat torus. 
The  non-geometry of the
background is then interpreted as the dependence of the
noncommutativity parameter on the base coordinate $x\in S^1$ of the
original torus
fibration,
with a monodromy around the circle that is a    Morita transformation.
Morita equivalence
 was
understood in~\cite{Seiberg1999} as the open string version of
T-duality in the decoupling limit, which is a symmetry of the noncommutative Yang--Mills theory. 

In the following we will re-examine D-branes on non-geometric spaces
in a more general setting, allowing for more general monodromies beyond the parabolic ones. 
The elliptic monodromies are of particular interest as they give 
string theory backgrounds directly, without the need to fibre over some base space~\cite{Dabholkar2002}.
For the
particular case of the $\IZ_4$ elliptic monodromy, we find that the low-energy
effective field theory is  defined on a non-geometric torus
and that all of its moduli, including the noncommutativity
parameter and the Yang--Mills coupling, have a monodromy in the base circle coordinate $x$ in such a way as to render the supersymmetric Yang--Mills
theory invariant under Morita duality. To the best of our knowledge,
such an example of a noncommutative fibration with non-geometric torus
fibres has not appeared before in the literature. The monodromy also interchanges
D0-brane and D2-brane charges, which swaps the roles of the rank of the
gauge fields and their topological charge in the worldvolume gauge theory.

We also study the effective
noncommutative gauge theories 
 in essentially doubled  spaces using 
the doubled twisted torus formalism
of~\cite{Hull:2007jy,DallAgata:2007egc,Hull2009,ReidEdwards2009}, in which D-branes have been
classified in~\cite{Albertsson2008}. Here we find a dependence of the noncommutative gauge theory on the dual
base coordinate $\wx\in S^1$, thus further
exemplifying the need of the doubled formalism in describing such configurations, another point of emphasis which is sometimes neglected
in the literature. The general picture of the effective theories on
D-branes in non-geometric polarisations of the doubled twisted torus
geometry is then that of a parameterised family of noncommutative
Yang--Mills theories with monodromies in $x$ or $\wx$ that are Morita
transformations. These   arise  as decoupling limits of backgrounds with monodromies
in $x$ or $\wx$ that are T-duality transformations.

An important feature of our considerations is the role of the doubled geometry.
For simple backgrounds, there is a conventional geometry which is seen by particles or momentum modes, while string winding modes will see a T-dual geometry.
 {However, in more complicated settings
there is a doubled geometry which cannot be disentangled to give a separate  geometry and dual geometry, and the momentum and  winding modes see different aspects of the full doubled geometry.
For a T-fold, there is a local split, referred to as  a polarisation in~\cite{Hull2004}, 
and local coordinates in  a 
patch can be split into spacetime coordinates and dual coordinates. However, globally this is not possible for a T-fold  as the T-duality transition functions mix the two kinds of coordinates so that there is no global polarisation.
For 
essentially doubled  spaces,
 the dependence of the background  on the coordinate conjugate to the winding number means that a conventional undoubled formulation is not possible even locally. 
Configurations  that are related to each other by  T-dualities all arise as different polarisations of   the same doubled geometry. For example, the four configurations in the duality chain~\eqref{eq:Tdualitychain}
 all arise as different polarisations of the same six-dimensional doubled space~\cite{Hull2009}. T-duality can be viewed as changing the polarisation~\cite{Hull2004}.
 
 A polarisation splits the doubled coordinates $\bX ^M$ into  ``spacetime coordinates'' $x^m$ and dual coordinates $\wx _m$.
 For a  conventional configuration, the background fields include the
 closed string metric $g$, the two-form $B$-field, and the dilaton
 $\phi$. These  background fields  depend only on $x^m$ and one obtains the usual spacetime interpretation, at least locally.
 For an essentially doubled configuration, some of the fields depend explicitly on the dual coordinates $\wx _m$.
For a conventional configuration with explicit dependence on
 a coordinate $x^\iota$, a generalised T-duality along 
 the vector field $\partial_\iota=\frac\partial{\partial x^\iota}$ will change the  dependence of the fields on $x^\iota$ to dependence of the fields on the dual coordinate $\wx _\iota$, resulting in an essentially doubled background. 
 
 The doubled geometry formulation of D-branes has some interesting features~\cite{Hull2004}.
 Consider a D$p$-brane wrapped on an $n$-torus $T^n$  with coordinates $x^m$, where $m=1,\dots, n$ and $p\le n$.
 Then the ends of open strings will have $p$  coordinates $x_{\rm
   D}^i$ satisfying Dirichlet boundary conditions on $T^n$ and $n-p$
 coordinates $x_{\rm N}^a$ satisfying Neumann boundary conditions.
 The doubled space is a torus $T^{2n}$ with coordinates $x^m, \wx_m$,
 with $m=1,\dots, n$. As T-duality interchanges Dirichlet and Neumann boundary conditions, 
 the coordinates dual to  $x_{\rm D}^i$ are $p$ coordinates  $\wx
 _{{\rm N}\,i}$ satisfying Neumann boundary conditions and the
 coordinates dual to $x_{\rm N}^a$ are $n-p$ coordinates 
$\wx _{{\rm D}\,a}$ with Dirichlet boundary conditions.
Then in the doubled torus there are precisely $n$ Dirichlet
coordinates $x_{\rm D}^i, \wx _{{\rm D}\,a}$, so that whatever the
value of $p$, the doubled picture is that of a D$n$-brane wrapping a maximally isotropic (Lagrangian) $n$-cycle in $T^{2n}$. As a result, a D$p$-brane is secretly a D$n$-brane in the doubled space.
 The polarisation determines  the subset of the $n$ Dirichlet
 directions which are regarded as physical, and changing the polarisation changes this subset: a T-duality that changes the polarisation from one with $p$ Dirichlet physical coordinates to one with $q$  Dirichlet physical coordinates is interpreted as taking a D$p$-brane to a D$q$-brane.
This picture was developed and extended to more general doubled spaces in \cite{Lawrence2006,Albertsson2008}.

The effective worldvolume theory on a D-brane is a  noncommutative
Yang--Mills theory coupling to a background open string metric $G_{\rm
  D}$ with noncommutativity bivector $\theta$ and gauge coupling
$g^{\phantom{\dag}}_{\rm YM}$.
In general the background fields $(G_{\rm D},\theta)$ as well as the
coupling $g^{\phantom{\dag}}_{\rm YM}$ can depend on the
coordinates $x^m$. The action  of T-duality on the closed string
background $(g,B,\phi)$ gives rise to Morita transformations of
$(G_{\rm D},\theta,g^{\phantom{\dag}}_{\rm YM})$, as we will review in \S\ref{sec:constantB},
 and   in our considerations of D-branes on non-geometric backgrounds
 we  find open string analogues of T-folds in which the dependence
 $(G_{\rm D}(x),\theta (x),g^{\phantom{\dag}}_{\rm YM}(x))$ on a circle coordinate $x$ can have 
  a monodromy that is a Morita transformation. Surprisingly, we also
  find open string analogues of essentially doubled backgrounds in
  which $(G_{\rm D},\theta,g^{\phantom{\dag}}_{\rm YM})$ have explicit dependence on a doubled coordinate
  $\wx$, possibly with a Morita monodromy. This suggests that the
  effective field theory should be defined on the full D$n$-brane in
  the doubled space, and
 so the fields can depend on  all  $n$ Dirichlet coordinates $x_{\rm
   D}^i,\wx _{{\rm D}\,a}$.}

One of the complications in the case of the flat three-torus with $H$-flux and its T-duals is that they do not define worldsheet conformal field theories, and so are not solutions of string theory.
However, there are string solutions in which  these appear as fibres.
The simplest case is that in which these are fibred over a line.
Taking an NS5-brane with transverse space $\mathbb{R}\times T^3$ and smearing over the 
$T^3$ gives a domain wall solution which is the product of six-dimensional Minkowski space with $\mathbb{R}\times T^3$, where there is constant $H$-flux over $T^3$ and the remaining fields depend explicitly on the coordinate of the transverse space $\mathbb{R}$~\cite{Hull:1998vy}.
T-duality then takes this to a metric on the product of $\mathbb{R}$ with the nilfold~\cite{Hull:1998vy,Lavrinenko:1997qa, Gibbons:1998ie,Lowe2003,Ellwood2006} 
 that is hyperk\" ahler, as was to be expected from the requirement  that the background is 
supersymmetric. Then further T-dualities in the chain  \eqref{eq:Tdualitychain} 
give T-folds and essentially doubled spaces fibred over a line; the proper incorporation of such spaces in string theory will be discussed further in \cite{Nipol}.
This leads to complications in the analysis of D-branes and decoupling limits in such backgrounds~\cite{Ellwood2006}. 

Due to the difficulties arising from such fibrations over a line or
other space, we will be particularly interested in examples that do
give string theory solutions directly without the need for introducing
a fibration. For the cases with elliptic monodromy, at special points
in the moduli space the background reduces to an orbifold defining a
conformal field theory and so provides a consistent string
background. However, we will also be interested in the elliptic
monodromy case at general points in the moduli space; these can arise
as fibres in which the moduli vary over a line or a higher-dimensional base space.

From the effective field theory point of view, the duality twisted reduction from ten dimensions gives a Scherk--Schwarz reduction of ten-dimensional supergravity to a seven-dimensional gauged supergravity. The fact that in general  the  product of the internal twisted torus with seven-dimensional Minkowski space does not define a conformal field theory, and so is not a supergravity solution, is reflected in the fact that the seven-dimensional supergravity has a scalar potential. In the parabolic monodromy case, the scalar potential has no critical points and so there are no Minkowski vacua, but there are domain wall solutions which lift to the ten-dimensional geometry given by the twisted torus fibred over a line.
In the elliptic monodromy case, there is a minimum of the potential corresponding to the orbifold compactification to seven dimensions~\cite{Dabholkar2002}, but again there are more general domain wall solutions in which the moduli vary over a line.

We will also consider the dilaton in what follows. For a given  background, T-duality will change the dilaton according to the Buscher rules.
Defining a conformal field theory requires the metric, $B$-field and dilaton to satisfy the beta-function equations, but it will be useful to consider general configurations of metric, $B$-field and dilaton without necessarily  requiring them to satisfy the beta-function equations -- they then define more general compactifications, as outlined above.

One of our motivations for revisiting these field theory perspectives
is to shed some light on the relevance of the noncommutative and nonassociative
deformations of closed string geometry which were recently purported to occur in
certain non-geometric
backgrounds~\cite{stnag1,Lust2010,Blumenhagen2011,Mylonas2012,Andriot2012}
(see e.g.~\cite{Szabo2018} for a review and further references). In contrast to these
analyses, here we work in a controlled setting with (doubled) twisted
tori and quantised fluxes, without any linear approximations and with
an exact effective field theory description of the string
geometry. Noncommutative and nonassociative geometries were suggested as global
(algebraic) descriptions of T-fold and $R$-flux  non-geometries respectively
in the mathematical framework of topological T-duality
in~\cite{Mathai2004,Bouwknegt2004,Brodzki2007,Bouwknegt2008}, which strictly
speaking only applies
to the worldvolumes of D-branes, but it was further suggested that such a
description should also apply to the closed string background
itself. Such a suggestion requires further clarification, insofar that in
closed string theory itself there is no immediate evidence for such nonassociative
structures. While we reproduce and generalise the noncommutative geometries on
D-branes in parabolic T-fold backgrounds, which were shown
by~\cite{Ellwood2006,Grange2006} to agree with the expectations from
topological T-duality, we do not directly find a nonassociative
geometry on D-branes in {$R$-folds. Instead, we find
that the decoupled noncommutative gauge theory on the D-branes
depends explicitly on the transverse doubled coordinate $\wx$, and so is essentially doubled and it appears that these cannot be fully understood in an undoubled space.}

The organisation of the  paper is as follows. In \S\ref{sec:constantB}
we briefly review, following~\cite{Seiberg1999}, the well-known description of the low-energy
effective dynamics of D-branes in constant $B$-fields in terms of
noncommutative gauge theory, and in particular its Morita duality on a torus
which is inherited from the T-duality symmetry of the closed string
background. In
\S\ref{sec:twistedtorus} we briefly review some general aspects of string theory compactifications on twisted tori, which are subsequently used to study the worldvolume gauge theories on D-branes
on three-dimensional T-folds via T-duality. We treat the cases of parabolic monodromies in \S\ref{sec:Tfoldparabolic} and of elliptic monodromies in \S\ref{sec:Tfoldelliptic}. We demonstrate, in both cases of parabolic and $\IZ_4$
elliptic twists, that there exist well-defined low-energy scaling
limits which completely decouple the open strings from closed strings,
and wherein the non-geometry of the T-fold background is manifested in
the open string sector as a parameterised family of noncommutative
gauge theories which are identified under Morita dualities determined
by the particular type of monodromy around $x\in S^1$. We give a physical interpretation
of the scaling limit which reproduces the mathematical description of
the field of noncommutative tori probed by the D-branes, and of the
Morita equivalence bimodules which implement the Morita duality
monodromies. In the
case of the elliptic monodromy we examine the theory at the orbifold
fixed point where we find that it is equivalent to an ordinary
commutative gauge theory on a flat torus for finite area and string
slope $\alpha'$. In \S\ref{sec:doubled} we review the doubled twisted
torus formalism and the classification of D-branes therein. This
setting allows us to take the final T-duality transformation that
describes D-branes in   essentially doubled  spaces, whose decoupled
worldvolume gauge theory is studied in~\S\ref{sec:Rfolds} where we
find that in this case the D-branes really probe a noncommutative
doubled geometry. For convenience, in Appendix~\ref{app:Buscher} we
summarise the Buscher T-duality rules including dimensionful factors.

\section{Open string dynamics in $B$-fields\label{sec:constantB}}

Consider the standard sigma-model for the embedding of an open string
worldsheet $\Sigma$ into a flat target space with constant metric $g$,
two-form gauge field  $B$ and dilaton $\phi$. We impose boundary conditions by
requiring that the boundary $\partial\Sigma$ is mapped to a
submanifold $\cW$ of spacetime, which is the worldvolume of a
D-brane. At tree-level in open string
perturbation theory, $\Sigma$ is a disk, which can be mapped to the
upper complex half-plane by a conformal transformation. 
The boundary of the upper   half-plane with coordinate $t \in\IR$ is 
then mapped to a curve $x^i(t)$ which is the worldline of the end of the string in the D-brane worldvolume $\cW$.
We are
interested in the dynamics of the open string ends
located on the D-brane. The two-point function of the $x^i$
on the boundary of the upper complex half-plane is given by~\cite{Seiberg1999}
\bea\label{eq:openprop}
\big\langle x^i(t)\, x^j(t')\big\rangle = -\alpha' \, G^{ij}
\log(t-t')^2 + \mbox{$\frac{\ii}2$} \, \Theta^{ij}\, {\rm sgn}(t-t') \ .
\eea
The metric $G$ and the bivector $\Theta$ determine the open
string geometry seen by the D-brane, and they are related to the
closed string metric $g$ and two-form $B$ by the open-closed string
relation
\bea
G^{-1}+\frac\Theta{2\pi\,\alpha'} := \big(g+2\pi\,\alpha'\,
B\big)^{-1} \ ,
\label{eq:openstringGTheta}\eea
which is equivalent to
\bea
G&=& g- (2\pi\,\alpha')^2\, B\, g^{-1}\, B \ , \nonumber\\[4pt]
\Theta&=& -(2\pi\,\alpha')^2\, (g+2\pi\,\alpha'\, B)^{-1}\, B\,
(g-2\pi\,\alpha'\, B)^{-1} \ .
\eea
Of particular interest is the second term in the open string
propagator \eqref{eq:openprop}, which depends only on the ordering of the
insertion points of open strings on the boundary of the disk and hence
leads to a well-defined target space quantity, independent of the
worldsheet coordinates.

In~\cite{Seiberg1999} it was shown that there is a consistent
decoupling limit where the string slope and closed string metric scale as $\alpha'=O(\epsilon^{1/2})$ and
$g_{ij}=O(\epsilon)$, with $\epsilon\to0$, which decouples the open
and closed string modes on the D-brane, and in which the bulk closed string geometry degenerates to a point. In this limit the first contribution to the propagator \eqref{eq:openprop} vanishes, while the open string
metric and bivector are finite and are given by
\bea
G_{\rm D}&=&-(2\pi\,\alpha')^2\, B\,g^{-1}\,B \ , \nonumber\\[4pt]
\theta&=&B^{-1} \ .
\eea
The open string interactions in scattering amplitudes
among tachyon vertex operators are captured in this limit by the
Moyal--Weyl star-product of fields $f,g$ on the D-brane worldvolume
given by
\bea\label{eq:MoyalWeyl}
f\star g = \boldsymbol\cdot\, \Big[\exp\big(\mbox{$\frac{\ii}2$}\,
\theta^{ij}\, \tfrac{\partial}{\partial
  x^i}\otimes\tfrac{\partial}{\partial x^j}\big)(f\otimes g)\Big] \ ,
\eea
where $\,\mbf\cdot\, (f\otimes g)= f\cdot g$ is the usual pointwise
multiplication of fields. The massless bosonic modes on the D-brane
are gauge and scalar fields whose low-energy dynamics in the
decoupling limit is described by noncommutative Yang--Mills
theory on $\cW$. The effective Yang--Mills coupling in the case of a D$p$-brane
gauge theory can be determined from the Dirac--Born--Infeld action and
is generally given by~\cite{Seiberg1999}
\bea
g_{\tiny\rm{YM}}^2 = \frac{(2\pi)^{p-2}}{(\alpha')^{(3-p)/2}} \ g_s \, \e^\phi \,
\bigg(\frac{\det(g+2\pi\,\alpha'\, B)}{\det g}\bigg)^{1/2} \ ,
\label{eq:gYMDp}\eea
where $g_s$ is the string coupling. This is finite in the decoupling
limit above if $g_s\,\e^\phi=O(\epsilon^{(3-p+r)/4})$, where $r$ is
the rank of the antisymmetric matrix $B$. These considerations can be extended to
curved backgrounds with non-constant $B$-field, including those with
non-vanishing $H$-flux $H=\dd B$~\cite{Cornalba2001,Herbst2001}, in
which case the Moyal--Weyl star-product \eqref{eq:MoyalWeyl} is
replaced by the more general Kontsevich star-product.

This story becomes particularly interesting in the case when
D$p$-branes wrap a $p$-dimensional torus $\cW=T^p$. In this case, T-duality of the
closed string background translates into open string T-duality which
acts on the D-brane charges. The T-duality group $O(p,p;\IZ)$ acts
on the closed string moduli
\bea
E=\frac1{\alpha'}\, \big(g+2\pi\,\alpha'\,B\big)
\eea
through the fractional linear transformations
\bea
\displaystyle{\widetilde E= (a\,E+b)\, (c\,E+d)^{-1}} \qquad
\mbox{for} \quad \begin{pmatrix} a & b \\ c &
    d \end{pmatrix} \in O(p,p;\IZ) \ .
\eea
The subgroup $SO(p,p;\IZ)$ is a proper symmetry of IIA or IIB string theory;
in the decoupling limit, this translates into $SO(p,p;\IZ)$ transformations of the
open string variables on the D$p$-brane given by
\bea
\widetilde G_{\rm D}&=&(c\,\theta+d)\,G_{\rm D} \,(c\,\theta+d)^\top \ , \nonumber \\[4pt]
\widetilde\theta&=&(a\,\theta+b)\, (c\,\theta+d)^{-1} \ , \nonumber
\\[4pt] \widetilde g^{\phantom{\dag}}_{\tiny\rm{YM}}&=&
g^{\phantom{\dag}}_{\tiny\rm{YM}}\, \big|\det(c\,\theta+d)\big|^{1/4}
\ .
\eea
The remarkable feature is that the noncommutative gauge theory on the
D$p$-brane inherits this T-duality symmetry. The transformation of the
bivector $\theta$ on its own is known from topological T-duality to
define a Morita equivalence between the corresponding noncommutative
tori $T_\theta^p$ and $T_{\widetilde\theta}^p$,  which mathematically preserves their K-theory
groups, or more physically the spectrum of D-brane charges on
$T_\theta^p$ and $T_{\widetilde\theta}^p$. Thus open string T-duality in the decoupling limit is a refinement of Morita
equivalence, which is referred to as Morita duality of noncommutative
gauge theory. 

In the mapping of T-duality of the closed string
background to Morita equivalence of noncommutative Yang--Mills theory
with gauge group $U(n)$,
it is generally necessary to 
introduce a closed two-form~\cite{Pioline1999,Seiberg1999,Ambjorn2000}
\bea
\Phi = \frac1{2n}\, Q_{ij}\, \dd x^i\wedge\dd x^j
\eea
on the D-brane worldvolume, which can be
thought of as an abelian background 't~Hooft magnetic flux, where
$Q_{ij}\in\IZ$ are the Chern numbers of a $U(n)$-bundle over $T^p$ of
constant curvature. 
The action is then constructed from a shifted form of the noncommutative field strength tensor
\bea\label{eq:FstarPhi}
\mathcal{F} = F_{\star} + \Phi \, \unit_n \, .
\eea
The dependence
on $\Phi$ simply serves  to  shift the classical vacuum of the
noncommutative gauge theory, giving the fields twisted periodic boundary
conditions around the cycles of $T^p$. Under T-duality, it is required
to transform as
\bea
\widetilde{\Phi} = (c\,\theta+d)\, \Phi\, (c\,\theta+d)^\top + c\,
(c\,\theta+d)^\top \ .
\eea
For example, if the components of the noncommutativity bivector
$\theta$ are rational-valued, then this can be used to provide a
Morita equivalence beween noncommutative Yang--Mills theory with
periodic gauge fields and ordinary Yang--Mills theory with gauge
fields having monodromies in $\IZ_n\subset U(n)$~\cite{Ambjorn2000}. The inclusion of
$\Phi$ also enables one to follow the T-duality orbits of
the charges of D-branes wrapping non-contractible cycles of even
codimension in the
D$p$-brane worldvolume, realised as topological charges in
the noncommutative gauge theory, which can be suitably arranged into
vectors of $SO(p,p;\IZ)$~\cite{Pioline1999,Seiberg1999}.

One purpose of this paper is to investigate this duality
in the cases of 
{twisted tori and the non-geometric backgrounds resulting from these under closed string
T-duality.}

\section{Compactification on the twisted torus\label{sec:twistedtorus}}

In the duality-twisted dimensional reductions of string theory on an {$n$-dimensional twisted torus} that we consider here, one first compactifies on an $n-1$-torus $T^{d}$, with $d=n-1$.
The   theory on this  internal space is then the conformal field theory with target space $T^{d}$, specified by a choice of modulus taking values in
the coset space $ O(d,d)/O(d)\times O(d)$,
which can be represented by a choice of metric $g$ and $B$-field on
the torus $T^{d}$. The group $O(d,d)$ then acts naturally on the combination
$E=\frac1{\alpha'}\, (g+2\pi\,\alpha'\, B)$ through fractional linear transformations. 
The automorphism group of the $T^{d}$ conformal field theory is the
subgroup $O(d,d;\IZ)$, 
which is  the T-duality group symmetry of the compactified string theory.
Configurations related by an $O(d,d;\IZ)$ transformation are physically equivalent, and so
the moduli space $ O(d,d)/O(d)\times O(d)$ should be identified under the action of $O(d,d;\IZ)$.

The next step is to compactify on a further circle $S^1$
and allow the modulus $E$ of the $T^{d}$ conformal field theory to depend on the  point $x\in S^1$. 
The $x$-dependence of $E(x)$ is determined by a map $\gamma:S^1\to O(d,d)$ given by
\bea\label{eq:gammaxM}
\gamma(x) = \exp(x\, M) \ ,
\eea
for a (dimensionless) mass matrix $M$ in the Lie algebra of $O(d,d)$. This map  has monodromy
\bea
\cM(\gamma)=\gamma(0)\, \gamma^{-1}(1) = \exp(M) \ .
\eea
For a consistent string theory background, this monodromy is required to be 
a symmetry of string theory, and so it must lie in the T-duality group
$O(d,d;\IZ)$~\cite{Hull:1998vy,Dabholkar2002}. The condition that this imposes on the mass matrix $M$ can be thought of as a  ``non-linear quantisation condition''.

The map $\gamma$ is a local section of a principal
bundle over $S^1$ with monodromy $\cM(\gamma)$. The
moduli of the theory depend on the coordinate $x$ through this
section, giving a parameterised family of conformal field theories over $S^1$ with
moduli $E(x)$, so that after a periodic shift $x\mapsto x+1$ around the
base $S^1$, the conformal field theory returns to itself up to the
monodromy $\cM(\gamma)$, which is an automorphism of the $T^{d}$ conformal field theory. Two such bundles are isomorphic, and hence define equivalent theories, if their monodromies lie in the same {$O(d,d;\IZ)$} conjugacy class.

Suppose that the monodromy $\cM(\gamma)$ takes values in the geometric
subgroup $GL(d,\IZ)$ of the duality group consisting of large
diffeomorphisms of the torus $T^{d}$. In this case we take $\gamma(x) $ in $GL(d,\IR)\subset
O(d,d)$, and represent $\gamma(x)$, $M$ and $\cM$ by $d\times d$ matrices. Then
compactifying on $T^{d}$, followed by compactification on $S^1$ with
the duality twist $\cM(\gamma)$ amounts to compactification on a torus
bundle $X$ over a circle $S^1$, often referred to as a twisted
torus. 
 We denote the local coordinates on this fibration by $(x,y^1,\dots,y^{d})$, where $x\in[0,1)$ is the coordinate on the base $S^1$ of radius $r$ and $(y^1,\dots,y^{d})\in[0,1)^{d}$ are coordinates on the fibres $T^{d}$. The metric is given by
\bea\label{eq:dsXd-1}
\dd s_X^2 = (2\pi\,r\, \dd x)^2+ h(\tau)_{ab} \, \dd y^a\, \dd y^b \ ,
\eea
where $h(\tau)$ is the metric on the $d$-torus, which depends on
moduli $\tau$ taking values in the coset space $GL(d,\IR)/SO(d)$. The
$x$-dependence of $\tau$ is defined by
\bea
h\big(\tau(x)\big)_{ab} = h(\tau^\circ)_{cd} \, \gamma(x)^c{}_a \,
\gamma(x)^d{}_b
\eea
for some fixed modulus $\tau^\circ$.

To determine the homologically stable cycles in $X$ which can be
wrapped by D-branes, it will prove useful to have another description
of these backgrounds. The twisted torus can also be described as the quotient
$G_\IZ\setminus G_\IR$ of an $n$-dimensional non-compact Lie group
$G_\IR$ by a cocompact discrete subgroup $G_\IZ$, so that much of the
local structure of the theory is the same as that for the
reduction on the group
manifold $G_\IR$. In particular, the left-invariant Maurer--Cartan
forms and the generators of the right action of $G_\IR$ are
well-defined on the compact space $G_\IZ\setminus G_\IR$. 

The generators $J_1,\dots,J_{d},J_x$ of the Lie algebra of $G_\IR$ then have brackets
\bea
[J_a,J_x] = M_a{}^b\, J_b \qquad \mbox{and} \qquad [J_a,J_b]=0 \ ,
\eea
where $M=(M_a{}^b)$ is {the $d\times d$ mass matrix satisfying
  $\gamma(x) = \exp(x\, M)$,} and $G_\IR$ may be described as a group
of $n\times n$ matrices
\bea
G_\IR = \Big\{\begin{pmatrix} \gamma^{-1}(x) & y \\ 0 & 1 \end{pmatrix} \ \Big| \ x,y^1,\dots,y^{d}\in\IR\Big\} \ .
\label{eq:GR}\eea
The left action of the discrete subgroup
\bea
G_\IZ = \Big\{\begin{pmatrix} \cM^{-\alpha} & \beta \\ 0 & 1 \end{pmatrix} \ \Big| \ \alpha,\beta^1,\dots,\beta^{d}\in\IZ\Big\} 
\eea
by multiplication on $G_\IR$ can be expressed in terms of the local coordinates as
\bea
x\longmapsto x+\alpha \qquad \mbox{and} \qquad y^a\longmapsto (\cM^{-\alpha})^a{}_b\, y^b+\beta^a \ ,
\eea
and the resulting quotient
\bea
X=G_\IZ\setminus G_\IR
\eea
is the required twisted torus construction. 

The $n$-manifold $X$ is
parallelisable, and the corresponding basis of left-invariant Maurer--Cartan forms is given by
\bea
\zeta^x= \dd x \qquad \mbox{and} \qquad \zeta^a=\gamma(x)^a{}_b\, \dd y^b \ .
\label{eq:MC1forms}\eea
They are globally defined one-forms on the torus bundle which obey the Maurer--Cartan equations
\bea
\dd\zeta^x=0 \qquad \mbox{and} \qquad \dd\zeta^a+M^a{}_b\, \zeta^x\wedge\zeta^b=0 \ .
\eea
The
metric \eqref{eq:dsXd-1} can then be rewritten as the
 left-invariant metric 
\bea\label{eq:mettt}
\dd s_X^2 = (2\pi\, r\, \zeta^x)^2 + h(\tau^\circ)_{ab} \, \zeta ^a\,
\zeta ^b \ .
\eea

\subsection{$SL(2,\IZ)$ monodromies}

In this paper we will study the examples with $n=3$ ($d=2$) in detail. In this case
the T-duality group of the string theory compactified on $T^2$ can be
factored as 
\bea
O(2,2;\IZ)\simeq \big(SL(2,\IZ)_\tau \times SL(2,\IZ)_\rho \big)
\rtimes \big( \IZ_2\times\IZ_2\big) \ .
\label{eq:dualitygroup}\eea
The first $SL(2,\IZ)$ factor is the mapping class
group of $T^2$ which acts geometrically by fractional linear transformations on the complex
structure modulus $\tau$ of $T^2$, while the second acts on the complexified
K\"ahler modulus $\rho$ whose imaginary part is the area of $T^2$ and
whose real part is the restriction of the two-form $B$-field to
$T^2$. 
The $\IZ_2\times\IZ_2$ factor can be taken to be generated by a reflection in one direction and a T-duality in one direction.

A further compactification on $S^1$ with the duality twist
$\cM(\gamma)$ in the geometric subgroup $SL(2,\IZ)_\tau$ is equivalent
to compactification on a $T^2$-bundle $X$ over $S^1$ with monodromy
$\cM(\gamma)$. The constant metric $ h(\tau)_{ab}$ on the $T^2$ fibers can be
written in terms of the complex structure modulus $\tau=\tau_1+\ii\tau_2 $ and  the
constant area modulus $A$ of the torus as
\bea
h(\tau)= \frac A {\tau_2}\,  \begin{pmatrix} 1& \tau_1\\ \tau_1& |\tau|^2 \end{pmatrix} \ .
\eea
The torus modulus transforms under $SL(2,\IZ)_\tau$ as $\tau\mapsto\cM[\tau]$ with
\bea
\cM[\tau]:=\frac{a\,\tau+b}{c\,\tau+d} \qquad \mbox{for} \quad \cM = \begin{pmatrix} a& b\\ c& d \end{pmatrix} \in SL(2,\IZ) \ .
\eea
In the $T^2$-bundle over $S^1$, the modulus varies with the circle
coordinate $x$ according to the $SL(2,\IR)$ transformation
\bea
\tau(x) = \gamma(x)[\tau^\circ]
\eea
for some fixed modulus $\tau^\circ$, so that
$\tau(x+1)=\cM\big[\tau(x)\big]$. The metric on the twisted
three-torus $X$ is given by \eqref{eq:dsXd-1}, which can be rewritten as
\bea
\dd s_X^2 = (2\pi\,r\, \dd x)^2+\frac A{\tau_2}\, \big|\dd y^1+ \tau\, \dd y^2\big|^2 \ .
\label{eq:dsX2general}\eea

In the following we will describe D-branes in  non-geometric
backgrounds associated with these
twisted three-tori.
For this, we wrap D$p$-branes around suitable $p$-cycles of
$X$ for $p=1,2$, which become D$(p+1)$-branes after T-duality to a T-fold background
characterised by a monodromy $\cM$ in a non-geometric subgroup of the
duality group $O(2,2;\IZ)$. We will study the corresponding noncommutative gauge theory on the D$(p+1)$-branes induced by the metric, $B$-field and dilaton of the T-fold background, in a scaling limit which decouples open and closed string modes. We shall generally find embeddings into non-geometric string theory 
of noncommutative Yang--Mills theory whose worldvolume geometry and
noncommutativity parameter vary over the base coordinate of a
non-geometric ``bundle'', and hence determine a parameterised family
of noncommutative gauge theories which is globally well-defined up to
Morita equivalence, the open string avatar of T-duality.

As conjugate monodromies define equivalent backgrounds $X$,  the monodromies leading to physically distinct configurations
 are classified by $SL(2,\IZ)$ conjugacy classes~\cite{Dabholkar2002}.
Following~\cite{Dabholkar2002,Dabholkar2005,Hull2005}, the conjugacy
classes can be classified into three sets: parabolic ($|{\rm
  Tr}(\cM)|=2$), elliptic ($|{\rm Tr}(\cM)|<2$) and
hyperbolic ($|{\rm Tr}(\cM)|>2$). In this paper we will concentrate on the examples of parabolic monodromies which generate integer translations
$\tau\mapsto\tau+m$ of the modular parameter $\tau$, and the $\IZ_4$
elliptic monodromies which generate the  inversion
$\tau\mapsto-\frac1\tau$. These examples capture many of the essential
features of the noncommutative gauge theories on D-branes in non-geometric
backgrounds. 

\section{D3-branes on T-folds: Parabolic monodromies\label{sec:Tfoldparabolic}}

The parabolic conjugacy classes of $SL(2,\IZ)$ generate monodromies
$\cM[\tau]=\tau+m$ of
infinite order and are labelled by an
integer $m\in\IZ$, with mass matrix $M$ and  monodromy matrix $\cM=\exp(M)$ where
\bea
\cM = \begin{pmatrix} 1 & m \\ 0 & 1 \end{pmatrix} \qquad \mbox{and}
\qquad M=\begin{pmatrix} 0 & m \\ 0 & 0\end{pmatrix} \ .
\label{eq:parmonodromy}\eea
The local section is given by
\bea
\gamma(x) = \begin{pmatrix} 1 & m\, x \\ 0 & 1 \end{pmatrix} \qquad
\mbox{with} \quad \tau(x) = \tau^\circ+m\, x \ ,
\label{eq:parabolictaux}\eea
where $\tau^\circ=\tau^\circ_1+\ii\tau^\circ_2$ is some constant
modulus, so that
\bea
\tau(0)=\tau^\circ \qquad \mbox{and} \qquad \tau(1)=\tau^\circ+m \ .
\eea
 The metric
can be brought to the form
\bea\label{eq:dsX}
\dd s_X^2 = (2\pi\, r\, \dd x)^2 + \frac A{\tau^\circ_2}\, \big(\dd
y^1+\omega\big)^2 + A\, \tau^\circ_2\, \big(\dd y^2\big)^2 \ ,
\eea
where $\omega:= (\tau^\circ_1+m\, x)\, \dd y^2$. This identifies the
twisted torus $X$ in this case as a circle bundle over $T^2$ of degree
$m$, with fibre coordinate $y^1$ and base coordinates $(x,y^2)$, while
$\omega$ is a connection on this bundle with Chern number $m$. The $B$-field vanishes and the dilaton is constant  in this background.

In this case
$G_\IR$ is the three-dimensional Heisenberg group whose generators
satisfy the Heisenberg algebra
\bea
[J_1,J_x]=m\, J_2 \qquad \mbox{and} \qquad [J_1,J_2]=0=[J_x,J_2]  \ .
\eea
Then the quotient by the discrete group action
\bea
x\longmapsto x+\alpha \ , \quad y^1\longmapsto y^1-\alpha\, m\,
y^2+\beta^1 \qquad \mbox{and} \qquad y^2\longmapsto y^2+\beta^2 \ 
\eea
for $\alpha,\beta^1,\beta^2\in\IZ$ is the three-dimensional
Heisenberg nilmanifold. A globally defined basis of one-forms on the
nilfold is given by
\bea
\zeta^x=\dd x \ , \quad \zeta^1 = \dd y^1+m\,x\,\dd y^2 \qquad
\mbox{and} \qquad \zeta^2 = \dd y^2 \ .
\eea
The Maurer--Cartan equations
\bea
\dd \zeta^x=0=\dd\zeta^2 \qquad \mbox{and} \qquad \dd\zeta^1 = m\,
\zeta^x\wedge \zeta^2
\eea
imply that $H^1(X,\IR)=\IR\oplus\IR$ is generated by $\zeta^x$ and
$\zeta^2$. By Poincar\'e duality, the second homology
$H_2(X,\IZ)=\IZ\oplus\IZ$ is   generated by the two-cycles
$\xi_{x,1}$ and $\xi_{1,2}$ dual to
$\zeta^x\wedge\zeta^1$ and $\zeta^1\wedge \zeta^2$, and in particular 
the two-cycle $\xi_{x,2}$ dual to $\zeta^x\wedge\zeta^2$ is
homologically trivial~\cite{Lawrence2006}. On the other
hand, from the Gysin sequence for $X$ viewed as a circle bundle it
follows that $H_1(X,\IZ)=\IZ\oplus\IZ\oplus\IZ_m$, where the
$\IZ$-valued classes are the one-cycles $\xi_x$ and $\xi_2$ 
dual to $\zeta^x$ and $\zeta^2$ on the $T^2$ base, while the
$\IZ_m$ torsion one-cycle $\xi_1$ is the class of the $y^1$ circle fiber.

This background is T-dual to a flat three-torus $T^3$ with $H$-flux: Applying the Buscher construction along the abelian
isometry generated by the global vector field $\frac\partial{\partial y^1}$ on
$X$ (see Appendix~\ref{app:Buscher}), T-duality maps the metric \eqref{eq:dsX} to the metric and $B$-field
\bea
\dd s_{T^3}^2 &=& (2\pi\,r \, \dd x)^2 + \frac{(2\pi \,\alpha')^2\, \tau^\circ_2}A\, \big(\dd y^1\big)^2
+ A\, \tau^\circ_2\,\big(\dd y^2\big)^2 \ , \nonumber \\[4pt] B_{T^3} &=& (\tau^\circ_1+m\,x) \
\dd y^1\wedge\dd y^2 \ .
\label{eq:T3gB}\eea
The $B$-field gives a constant $H$-flux
\bea\label{eq:HT3}
H_{T^3}=\dd B_{T^3}= m\,\dd x\wedge\dd
y^1\wedge\dd y^2
\eea
on $T^3$.
This has a monodromy in $SL(2,\IZ)_\rho$,
\bea\cM[\rho]=\rho+m \ ,
\eea
giving a shift in $B_{T^3}$ by  $m\, \dd y^1\wedge\dd y^2$, which represents an integral cohomology class.

\subsection{Worldvolume geometry}

Let us now wrap a D2-brane around the non-trivial two-cycle
$\xi_{x,1}$. T-duality in the $y^1$-direction then maps the D2-brane to a D1-brane wrapped around the one-cycle
dual to $\zeta^x$ in the flat three-torus $T^3$ with metric and $B$-field
given by \eqref{eq:T3gB}, and constant $H$-flux \eqref{eq:HT3}. These are both allowed D-brane configurations, according to  the doubled torus analysis of~\cite{Lawrence2006}.

On the other hand, we can consider T-duality along the abelian
(covering space) isometry generated by the vector field
$\frac\partial{\partial y^2}$ which maps the D2-brane to a D3-brane filling the T-fold.
The metric and $B$-field are given by
\bea
g &=& (2\pi\,r\, \dd x)^2 + \frac{\tau_2(x)}{|\tau(x)|^2}\,
\Big(A \, \big(\dd y^1\big)^2+\frac{(2\pi \, \alpha')^2}{A}\,
\big(\dd y^2\big)^2 \Big) \ , \nonumber \\[4pt]
B &=& \frac{\tau_1(x)}{|\tau(x)|^2} \ \dd y^1\wedge \dd y^2 \ ,
\label{eq:TfoldgB}\eea
together with the dilaton field
\bea
\e^{\phi(x)} = \bigg(\frac{2\pi\,\alpha'}A\, \frac{\tau_2(x)}{|\tau(x)|^2}
\bigg)^{1/2} \ .
\label{eq:Tfolddilatonparabolic}\eea
where
\bea
\tau(x)=\tau^\circ_1+m\, x + \ii \tau^\circ_2 \, .
\eea
The area
of the $T^2$ fibres with coordinates $(y^1,y^2)$ is 
\bea 
{\cal {A} }=   2\pi \, \alpha'\, \frac{\tau_2(x)}{|\tau(x)|^2} \ .
\eea
Then 
 the K\"ahler modulus of the $T^2$ fibres  is
\bea 
\rho :=  B_{12} +\frac  \ii {2\pi \, \alpha'} \,{\cal {A} }
=
\frac{\tau_1(x)+\ii \tau_2(x)}{|\tau(x)|^2}
=
 \frac 1 {\bar \tau (x)}
\eea
so that, as $\tau (x+1)=\tau (x)+m$,
this is a T-fold with monodromy
\bea \rho (x+1) =   \frac {\rho (x)} {1+m\,\rho (x)}
\eea
 in $SL(2,\IZ)_\rho$.

Let us now transform to the open string variables seen by the
D3-brane~\cite{Seiberg1999,Cornalba2001}. These are the open string metric $G$ and
noncommutativity bivector $\Theta$ defined from \eqref{eq:TfoldgB} through
\eqref{eq:openstringGTheta}.
Explicit calculation from \eqref{eq:TfoldgB} gives a worldvolume
$\cW_{\rm D3}$ 
with the topology of $S^1\times T^2$ and
\bea
G &=& (2\pi\,r \, \dd x)^2 + \frac A{\tau^\circ_2} \, \big(\dd y^1\big)^2
+ \frac{(2\pi \,\alpha')^2}{A\, \tau^\circ_2}\,\big(\dd y^2\big)^2 \ , \nonumber \\[4pt] \Theta &=& (\tau^\circ_1+m\,x) \,
\frac\partial{\partial y^1} \wedge\frac\partial{\partial y^2} \ .
\label{eq:openstringparabolic}\eea

\subsection{Noncommutative Yang--Mills theory}

In order to get a low-energy limit with pure gauge theory on $\cW_{\rm D3}$
in which
the massive string modes are decoupled and gravity is non-dynamical, we need to take the zero slope limit $\alpha'\to0$ while keeping $G$ and $\Theta$ fixed, which in the present case means keeping the parameters
\bea
r_1:=\bigg(\frac A{4\pi^2\, \tau_2^\circ}\bigg)^{1/2} \qquad
\mbox{and} \qquad r_2:= \frac{\alpha'}{(A\, \tau_2^\circ)^{1/2}}
\label{eq:scalingradii}\eea
fixed. This can be achieved by the scaling limit $\alpha'=O(\epsilon^{1/2})$, $A=O(\epsilon^{1/2})$ and $\tau_2^\circ=O(\epsilon^{1/2})$ with $\epsilon\to0$, and with all other parameters, including the $B$-field, held fixed. In this limit the closed string metric from \eqref{eq:TfoldgB} degenerates along $T^2$, taking the area to zero while fixing $B$, whereas the open string parameters on the D3-brane become
\bea
\dd s_{\rm D3}^2 &=& (2\pi\,r \, \dd x)^2 + \big(2\pi\,r_1 \, \dd y^1\big)^2
+ \big(2\pi\,r_2 \, \dd y^2\big)^2 \ , \nonumber \\[4pt] \theta &=& (\tau^\circ_1+m\,x) \,
\frac\partial{\partial y^1} \wedge\frac\partial{\partial y^2} \ ,
\label{eq:dsD3parabolic}\eea
where in the particular instance of a parabolic twist the open string bivector $\Theta$ from \eqref{eq:openstringparabolic} and its zero slope limit $\theta$ happen to coincide.

Finally, the effective Yang--Mills coupling can be determined from
\eqref{eq:gYMDp}, which in the present case with $p=3$ is the constant
\bea
g_{\tiny\rm{YM}}^2=
\bigg(\frac{(2\pi)^3\, \alpha' \, g^2_s}{A\,
  \tau^\circ_2}\bigg)^{1/2} \ .
\eea
In order to obtain a well-defined quantum gauge theory, we thus require that
\bea
\bar g_s^2:= \frac{2\pi\,\alpha'\, g^2_s}{A\, \tau^\circ_2}
\eea
remains finite in the zero slope limit, which implies that the string
coupling scales as $g_s=O(\epsilon^{1/4})$, which is consistent with the
perturbative regime of the string theory that we are working in. Then
\bea\label{eq:gYMparabolic}
g_{\rm{\tiny YM}}^2=2\pi\,\bar g_s
\eea
is indeed finite in the limit $\alpha'\to0$.

Since $\frac\partial{\partial y^a}\theta=0$, the supersymmetric noncommutative Yang--Mills theory on the D3-brane
is defined by multiplying fields $f,g$ on $S^1\times T^2$ together with the
Kontsevich star-product~\cite{Lowe2003} (see also~\cite{Hannabuss2009})
\bea
f\star g = \, \mbf\cdot\, \Big[\exp\big(\, \mbox{$\frac\ii2$} \, \theta(x)\,
(\mbox{$\frac\partial{\partial y^1} \otimes \frac\partial{\partial y^2} - \frac\partial{\partial y^2}
\otimes \frac\partial{\partial y^1})$} \, \big) (f\otimes g)\Big] \ .
\label{eq:starparabolic}\eea
Defining $[f,g]_{\star}:=f\star g-g\star f$, this gives a quantisation of the three-dimensional Heisenberg algebra
\bea
[y^1,y^2]_{\star} = \ii\theta(x) \qquad \mbox{and} \qquad
[y^1,x]_{\star}=0= [y^2,x]_{\star} \ .
\label{eq:starHeisenberg}\eea
For fixed
$x\in S^1$ the star-product \eqref{eq:starparabolic} defines a
noncommutative torus $T^2_{\theta(x)}$, which
means geometrically that 
varying $x\in S^1$ determines a field of noncommutative tori in
the D3-brane worldvolume $\cW_{\rm
  D3}$~\cite{Mathai2004,Hannabuss2009}. 

The noncommutative torus $T^2_{\theta(x)}$ has Morita equivalence group
\bea
SO(2,2;\IZ) \simeq \big( SL(2,\IZ)_{\theta}\times SL(2,\IZ)_\tau \big) / \IZ _2  \ .
\eea 
(The $ \IZ _2$ factor is generated by $(-\unit,-\unit)\in SL(2,\IZ)\times SL(2,\IZ)$.)
Morita transformations in this group leave invariant the noncommutative
gauge theory on the D3-brane filling the T-fold compactification. Here
the roles of the two $SL(2,\IZ)$ factors in the original duality group
\eqref{eq:dualitygroup} are interchanged: The  $SL(2,\IZ)_\rho$
factor in \eqref{eq:dualitygroup} now acts on the D3-brane torus 
with  \eqref{eq:dsD3parabolic}
as the geometric action of 
 the mapping class group $SL(2,\IZ)_\tau$ of the $T^2$ \lq fibres'  with coordinates $(y^1,y^2)$, 
  leaving $\theta$, the Yang--Mills coupling $g^{\phantom{\dag}}_{\tiny\rm{YM}}$ and the area $V=4\pi^2\,r_1\,r_2$ of the torus $T^2$ with the metric $\dd s_{\rm D3}^2$ unchanged, while 
the   $SL(2,\IZ)_\tau$ factor in \eqref{eq:dualitygroup}  now acts
as the $SL(2,\IZ)_\theta$ Morita  transformations
\bea
\cM[V]&=&V\, (c\,\theta+d)^2 \ , \nonumber \\[4pt]
\cM[\theta]&=&\frac{a\,\theta+b}{c\,\theta+d} \ , \nonumber \\[4pt]
\cM[g^{\phantom{\dag}}_{\tiny\rm{YM}}]&=& g^{\phantom{\dag}}_{\tiny\rm{YM}}\, | c\,\theta+d | ^{1/2} \ .
\label{eq:SL2Ztheta}\eea
This is   the old statement~\cite{Seiberg1999} that Morita
equivalence is precisely the structure inherited from T-duality in the
decoupling limit.

Thus by wrapping a D3-brane we gain an alternative perspective on the
non-geometric nature of the T-fold background in terms of
noncommutative gauge theory: under a monodromy around the circle coordinate $x$, the noncommutativity parameter transforms as $\theta(x+1)=\theta(x)+m$, which is precisely an $SL(2,\IZ)_\theta$ Morita transformation by the pertinent monodromy matrix \eqref{eq:parmonodromy}
\bea
\theta(x+1) = \cM\big[\theta(x)\big] \ .
\eea
The identification of monodromies in $x$ via T-duality in the closed string
sector is realised in the open string sector via Morita equivalence in
Yang--Mills theory on a noncommutative torus.  Under the parabolic monodromy, all other parameters of the gauge
theory, including the open string metric $\dd s_{\rm D3}^2$, the area $V$ and the
Yang--Mills coupling constant $g^{\phantom{\dag}}_{\tiny\rm{YM}}$, are
invariant. 

To summarise, in the case of parabolic twists we have found that
although closed strings see non-geometry, open strings see an
undeformed conventional geometric torus but the original closed string
non-geometry is now reflected in the noncommutativity bivector $\theta$ in the dual
gauge theory description of D3-branes as a $\theta$-deformed
noncommutative supersymmetric Yang--Mills theory. The T-duality monodromy for
the geometric moduli of the closed string geometry is mapped to a
Morita monodromy for the moduli of noncommutative Yang--Mills theory.

\subsection{Interpretation of the decoupling limit\label{sec:interpretation}}

We can give a physical derivation of this noncommutative gauge theory
by adapting the description of~\cite{Douglas1997} which considered the case of
vanishing fluxes and constant $B$-field. The essential features can
be seen already in the low-energy effective theory of a D1-brane on the
twisted torus $X$ wrapping the torsion one-cycle $\xi_{1}$, and placed at
$y^2=0$ and any fixed point $x\in S^1$.
We can think of the original torus fibres $T^2$ of $X$ as the complex plane
$\IC$, with coordinate $z=y^1+\ii y^2$, quotiented by the translations
$z\mapsto z+\alpha$ and $z\mapsto z+\beta\,\tau(x)$ for
$\alpha,\beta\in\IZ$. In the scaling limit $\tau_2^\circ\to0$ taken
above, the torus fibre degenerates to the flat cylinder $S^1\times\IR$
with coordinate $y^1\in[0,1)$ quotiented by the additional
translations $y^1\mapsto y^1+\beta\,\theta(x)$; this is not a
conventional Hausdorff space for generic values of $x\in S^1$, but can
be precisely interpreted as the noncommutative torus
$T_{\theta(x)}^2$, which for irrational values of $\theta(x)$ is
sometimes called the `irrational rotation algebra'. In this geometric
picture, the Morita invariance under parabolic monodromies around the base
circle is trivially realised as the equality $T^2_{\theta(x+1)} =
T^2_{\theta(x)+m}=T^2_{\theta(x)}$ under the identification
of the periodic coordinate $y^1$ with $y^1+m$.

In the gauge theory on the D1-brane, there are additional light states
formed by strings winding $w^2$ times around $y^2$, viewed as open
strings connecting the D1-brane and its images on the covering space
over $y^2$, which have mass proportional to $w^2\,\tau_2^\circ$. The
complete low-energy spectrum for $\tau_2^\circ\to0$ is thus obtained
by considering fields $f_{w^2}(y^1)$ with an arbitrary dependence on
both $y^1\in[0,1)$ and on $w^2\in\IZ$. The open string starting at $(y^1,0)$ ends at
$(y^1,w^2\,\tau_2^\circ)$, which is identified with the point
$(y^1-w^2\,\theta(x),0)$ on the twisted torus. Since open strings
interact via concatenation of paths, in $(y^1,w^2)$ space the
interaction of two fields $f_{w^2}$ and $\tilde f_{\tilde w^2}$ is given by
\bea
f_{w^2}\big(y^1\big)\,\tilde f_{\tilde w^2}\big(y^1-w^2\,\theta(x)\big) = f_{w^2}\big(y^1\big)\, \exp\big(-w^2\, \theta(x)\, \mbox{$\frac\partial{\partial y^1}$}\big)\, \tilde f_{\tilde w^2}\big(y^1\big) \ .
\eea
By T-duality along the vector field $\frac\partial{\partial y^2}$, which maps the
winding number $w^2$ to a momentum mode $p_2$, followed by the usual
Fourier transform of $p_2=-\ii\frac\partial{\partial y^2}$ to $y^2$, this interaction is given by the
noncommutative star-product $f\star\tilde f$ from
\eqref{eq:starparabolic} in the gauge theory on the D2-brane in the
dual T-fold frame; in particular, this shows that the star-product
\eqref{eq:starparabolic} is invariant under the monodromy $\theta(x+1)=\theta(x)+m$. The open string metric on the D2-brane in the
scaling limit is obtained  from \eqref{eq:dsD3parabolic} with all other parameters as above
and is the metric on a flat square torus with radii $r_1,r_2$. 
 This gives a family of D2-brane
gauge theories on $T^2_{\theta(x)}$ parameterised by $x\in S^1$, such
that after a monodromy $x\mapsto x+1$ the noncommutative gauge
theory returns to itself up to Morita equivalence, which is a symmetry of the theory; in particular, this
leaves the noncommutative Yang--Mills action $S_x^{\tiny\rm{YM}}$
invariant: $S_{x+1}^{\tiny\rm{YM}} = S_x^{\tiny\rm{YM}}$. The fibre
over $x$ of this
parameterised family of noncommutative gauge theories is dual to the
low-energy
effective theory of a D0-brane placed at $y^1=y^2=0$ and $x\in S^1$ on the three-torus $T^3$ with constant
$H$-flux~\eqref{eq:HT3}.

\section{D2-branes on T-folds: Elliptic monodromies\label{sec:Tfoldelliptic}}

The elliptic conjugacy classes of $SL(2,\IR)$ are   
matrices $\cM=\exp(M)$ that are  conjugate to rotations, so that they are of the form 
\bea
\cM= U \, \begin{pmatrix} \cos(m\,\vartheta) & \sin(m\,\vartheta) \\ -\sin(m\,\vartheta) & \cos(m\,\vartheta) \end{pmatrix} \, U^{-1} \qquad \mbox{and} \qquad M = U\, \begin{pmatrix} 0& m\,\vartheta \\ -m\,\vartheta & 0 \end{pmatrix} \, U^{-1} \ ,
\label{eq:ellipticmonodromy}\eea
where $U\in SL(2,\IR)$, the angle $\vartheta\in(0,\pi]$ and $m\in\IZ$.
The elliptic conjugacy classes of $SL(2,\IZ)$ are    matrices of integers that are in elliptic conjugacy classes of $SL(2,\IR)$. This is highly restrictive, and the only   angles for which there is a $U$ such that $\cM$ is integer-valued are
$\vartheta=\pi, \frac{2\pi}3, \frac\pi2, \frac\pi3$. These give
matrices of finite order, generating the cyclic groups
$\IZ_2,\IZ_3,\IZ_4,\IZ_6$ respectively, which provide the four possible choices of elliptic monodromies.

For $\vartheta=\pi$  (and $m\in2\,\IZ+1$) and  $\vartheta=\frac\pi2$ (and $m\in4\,\IZ+1$) 
  the required conjugation is trivial, $U=\unit$. These $SL(2,\IZ)$ transformations then act on $\tau $ by 
 $\cM[\tau]=\tau$
and $\cM[\tau]=-\frac1\tau$ respectively.
For $\vartheta=\frac{2\pi}3$ and $\vartheta=\frac\pi3$ (with $m=1$),
the conjugation matrix is
\bea
U = \sqrt{\mbox{$\frac2{\sqrt3}$}} \, \bigg( \begin{matrix}
1 & \frac12 \\ 0 & \frac{\sqrt3}2 \end{matrix} \bigg) \ .
\eea
These generate
$\cM[\tau]=-\frac1{\tau+1}$
and  $\cM[\tau]=-\frac{\tau+1}\tau$ respectively. 

The
local section is given by
\bea
\gamma(x) = U \, \begin{pmatrix} \cos(m\,\vartheta\, x) & \sin(m\,\vartheta\,x) \\ -\sin(m\,\vartheta\,x) & \cos(m\,\vartheta\,x) \end{pmatrix} \, U^{-1} \ .
\eea
For definiteness, we now focus 
 our discussion on the case 
of $\IZ_4$ monodromies with $U=\unit$,
$\vartheta=\frac\pi2$ and $m\in4\,\IZ+1$ so that
\bea
\cM=  
\begin{pmatrix} 0 & 1 \\ -1 & 0 \end{pmatrix} \ .
  \label{eq:Z4monodromy}
\eea
 Then the complex structure modulus is
\bea
\tau(x) = \frac{\tau^\circ\cos(m\,\vartheta\,x)+\sin(m\,\vartheta\,x)}{-\tau^\circ\sin(m\,\vartheta\,x)+\cos(m\,\vartheta\,x)} \ ,
\label{eq:elliptictaux}\eea
with
\bea 
\tau(0)=\tau^\circ \qquad \mbox{and} \qquad \tau(1) = -  \frac 1
{\tau^\circ} \ .
\eea

In this case $G_\IR=ISO(2)$ is the isometry group of the Euclidean plane $\IR^2$ whose generators satisfy
\bea
[J_1,J_x]=m\,\vartheta\, J_2 \ , \qquad [J_2,J_x]=-m\,\vartheta\, J_1 \qquad \mbox{and} \qquad [J_1,J_2]=0 \ .
\eea
The group manifold of $ISO(2)$ has the topology of $S^1\times\IR^2$ which is compactified by the discrete group action
\bea
 y^a \longmapsto  y^a+\beta^a \ ,
\eea
where $\beta^a\in\IZ$ for
$a=1,2$; then $X$ is topologically $S^1\times T^2$. For $U=\unit$, the Maurer--Cartan equations
\bea
\dd\zeta^x=0 \ , \qquad \dd\zeta^1=-m\,\vartheta\, \zeta^x\wedge\zeta^2 \qquad \mbox{and} \qquad \dd\zeta^2=m\,\vartheta\,\zeta^x\wedge\zeta^1
\eea
imply that $H^1(X,\IR)=\IR$ is generated by $\zeta^x$. By Poincar\'e
duality it follows that $H_2(X,\IZ)=\IZ$ is generated by $\xi_{1,2}$,
and in particular now both $\xi_{x,1}$ and $\xi_{x,2}$ are
homologically trivial two-cycles. On the other hand, for the $\IZ_4$ monodromy,
$H_1(X,\IZ)=\IZ\oplus\IZ_2$ is generated by the $\IZ$-valued $S^1$ base one-cycle $\xi_x$
dual to $\zeta^x$, with the $\IZ_2$ torsion one-cycle $\xi_1$ given by the class of the $y^1$ circle fibre~\cite{Lawrence2006}.

The metric is given by \eqref{eq:mettt}. For the parabolic monodromy, T-dualising in $y^1$ gave a $T^3$ with $H$-flux while T-dualising in $y^2$ gave a T-fold, but for this elliptic case, dualising in either $y^1$ or $y^2$ gives the same result, which is a T-fold with $H$-flux. 
Starting with the twisted torus metric
\eqref{eq:dsX2general}, we apply the Buscher construction along the
abelian isometry generated by the vector field $\frac\partial{\partial y^2}$ to get  a non-geometric background with metric and $B$-field given   by
\bea
g &=& (2\pi\,r\, \dd x)^2 + \frac{\tau_2(x)}{|\tau(x)|^2}\,
\Big(A \, \big(\dd y^1\big)^2+\frac{(2\pi \, \alpha')^2}{A}\,
\big(\dd y^2\big)^2 \Big) \ , \nonumber \\[4pt]
B &=& \frac{\tau_1(x)}{|\tau(x)|^2} \ \dd y^1\wedge \dd y^2 \ ,
\label{eq:TfoldgBelliptic}\eea
together with the dilaton field
\bea
\e^{\phi(x)} = \bigg(\frac{2\pi\,\alpha'}A\, \frac{\tau_2(x)}{|\tau(x)|^2}
\bigg)^{1/2} \ .
\label{eq:Tfolddilatonelliptic}\eea
Here the K\"ahler modulus of the $T^2$ fibres with coordinates $(y^1,y^2)$ is
\bea 
\rho(x) = \frac 1 {\bar \tau(x)}
\eea
so that this is a T-fold with monodromy
\bea \rho (x+1) = -  \frac 1 {\rho (x)}
\eea
 in $SL(2,\IZ)_\rho$.

\subsection{Worldvolume geometry}

Unlike the parabolic case, here we cannot wrap a D2-brane on the base
$S^1$ of the twisted torus. Moreover, unlike the parabolic case, T-dualising the twisted torus with
$\IZ_4$ elliptic monodromy in either of the torus fibre directions results in something non-geometric.
 Instead, we wrap a D1-brane around the
torsion one-cycle $\xi_1$ as we did in \S\ref{sec:interpretation}.
T-dualising   $y^2$ gives  a D2-brane in the  T-fold background 
with metric and $B$-field in
\eqref{eq:TfoldgBelliptic},
and dilaton in
\eqref{eq:Tfolddilatonelliptic}.
Transforming now to the open string metric and noncommutativity bivector on the D2-brane using \eqref{eq:openstringGTheta} we find
\bea
G &=& \frac A{\tau_2(x)} \, \big(\dd y^1\big)^2
+ \frac{(2\pi \,\alpha')^2}{A\, \tau_2(x)}\,\big(\dd y^2\big)^2 \ , \nonumber \\[4pt] \Theta &=& \tau_1(x) \,
\frac\partial{\partial y^1} \wedge\frac\partial{\partial y^2} \ .
\label{eq:openstringelliptic}\eea

\subsection{Noncommutative Yang--Mills theory\label{sec:NCYMelliptic}}

In the zero slope limit with the radii \eqref{eq:scalingradii} held fixed, the closed string metric from \eqref{eq:TfoldgBelliptic} is again degenerate, while the decoupled open string noncommutative geometry is described by the metric and bivector
\bea
\dd s_{\rm D2}^2 &=& \big(-\tau_1^\circ
\sin(m\,\vartheta\, x)+ \cos(m\,\vartheta\,x)\big)^2\, \Big( (2\pi\,r_1)^2 \, \big(\dd y^1\big)^2
+ (2\pi\,r_2)^2 \,\big(\dd y^2\big)^2\Big) \ , \nonumber \\[4pt] \theta &=& \frac{\tau_1^\circ\cos(m\,\vartheta\,x)+\sin(m\,\vartheta\,x)}{-\tau_1^\circ\sin(m\,\vartheta\,x)+\cos(m\,\vartheta\,x)} \
\frac\partial{\partial y^1} \wedge\frac\partial{\partial y^2} \ .
\label{eq:dsD2elliptic}\eea
Again since $\frac\partial{\partial y^a}\theta=0$, the star-product incorporating the dynamics of open strings in this background is given in the same
form \eqref{eq:starparabolic} quantising the three-dimensional algebra
\eqref{eq:starHeisenberg}, which however is no longer based on a Lie
algebra but rather some quantum deformation of the Heisenberg Lie
algebra determined by the
discrete parameters $m\in4\,\IZ+1$ and $\vartheta=\frac\pi2$.

As before, the non-geometric nature of the closed string background is captured in
the noncommutative gauge theory on the D2-brane via Morita
equivalence. Under a monodromy in the circle coordinate $x$, the
noncommutativity parameter transforms in the expected way from
\eqref{eq:SL2Ztheta} under an
$SL(2,\IZ)_\theta$ Morita transformation corresponding to the elliptic
monodromy matrix \eqref{eq:ellipticmonodromy}:
\bea
\theta(x+1) = \frac{\cos(m\,\vartheta)\ \theta(x) +
  \sin(m\,\vartheta)}{-\sin(m\,\vartheta)\ \theta(x) +
  \cos(m\,\vartheta)} = \cM\big[\theta(x)\big] \ .
\eea
For 
$m\in4\,\IZ+1$ and $\vartheta=\frac\pi2$
this reduces to
\bea
\theta(x+1)=- \frac 1{\theta (x)} \ .
\eea
However, {in contrast with the case  of parabolic twists, here} the metric on the
D2-brane worldvolume $\cW_{\rm D2}$ is not globally well-defined, so
that the open string sector now simultaneously probes both a non-geometric
and a noncommutative space. This is exactly what is needed to compensate the Morita equivalence
of the corresponding noncommutative fibre tori $T^2_{\theta(x)}$ and
render the noncommutative Yang--Mills theory on $\cW_{\rm D2}$
invariant; in particular, the area of the non-geometric worldvolume
\bea
V(x) = 4\pi^2\, r_1\, r_2\, \big(-\tau_1^\circ
\sin(m\,\vartheta\, x)+ \cos(m\,\vartheta\,x)\big)^2
\eea
transforms under a monodromy $x\mapsto x+1$ in the expected way from
\eqref{eq:SL2Ztheta} under the Morita duality corresponding to
\eqref{eq:ellipticmonodromy}:
\bea
V(x+1) = V(x) \ \big(-\sin(m\,\vartheta)\ \theta(x) +
  \cos(m\,\vartheta) \big)^2 = \cM\big[V(x)\big] \ .
\eea
For 
$m\in4\,\IZ+1$ and $\vartheta=\frac\pi2$
this reduces to
\bea
V(x+1) = V(x) \ \theta(x)^2  \ .
\eea

Finally, the Yang--Mills coupling of the decoupled noncommutative
gauge theory in the non-geometric $T^2$-bundle over $S^1$ is
$x$-dependent and is computed from \eqref{eq:gYMDp} with $p=2$ to get
\bea
g^{\phantom{\dag}}_{\tiny\rm{YM}}(x)^2=\bigg(\frac{2\pi \,g^2_s}{A\, \tau_2(x)} \bigg)^{1/2} \ .
\eea
In this case, it is the combination
\bea
\bar g_s^2 := \frac{g_s^2}{2\pi \, A \, \tau_2^\circ}
\eea
which must be fixed in the zero slope limit, so that now the string
coupling scales as $g_s=O(\epsilon^{1/2})$. Then the Yang--Mills
coupling in the zero slope limit is still $x$-dependent and given by 
\bea
g^{\phantom{\dag}}_{\tiny\rm{YM}}(x)^2= 2\pi\,\bar g_s\, \big| -\tau_1^\circ
\sin(m\,\vartheta\, x)+ \cos(m\,\vartheta\,x) \big| \ .
\label{eq:gYMelliptic}\eea
Hence the Yang--Mills coupling also transforms in the expected way
from \eqref{eq:SL2Ztheta} under the Morita duality corresponding to
\eqref{eq:ellipticmonodromy}:
\bea
g^{\phantom{\dag}}_{\tiny\rm{YM}}(x+1) = g^{\phantom{\dag}}_{\tiny\rm{YM}}(x) \ \big|-\sin(m\,\vartheta)\ \theta(x) +
  \cos(m\,\vartheta) \big|^{1/2} = \cM\big[g^{\phantom{\dag}}_{\tiny\rm{YM}}(x)\big] \ .
\eea
For 
$m\in4\,\IZ+1$ and $\vartheta=\frac\pi2$
this is
\bea
g^{\phantom{\dag}}_{\tiny\rm{YM}}(x+1) = g^{\phantom{\dag}}_{\tiny\rm{YM}}(x) \ \big| \theta(x)   \big|^{1/2} 
  \ .
\eea

Thus again we obtain a family of noncommutative D2-brane gauge
theories on $T_{\theta(x)}^2$ parameterised by $x\in S^1$. We stress that there is a well-defined action of the Morita duality transformations on the gauge theory: all three moduli -- the area, the noncommutativity parameter and the gauge coupling -- transform in a way that conspires to leave the gauge theory invariant. The
underlying noncommutative geometries here are new and generalise the field
of noncommutative tori obtained previously in the case of parabolic
monodromies. 

\subsection{Interpretation of the Morita equivalence monodromy\label{sec:Morita}}

The case of
elliptic mondromies exhibits another new feature compared to the case
of parabolic twists. Recall that the parabolic monodromy 
 affects
only the noncommutativity bivector $\theta(x)$ and is an exact
invariance of the noncommutative torus at the topological level,
$T^2_{\theta(x+1)}=T^2_{\theta(x)}$; in particular the
star-product \eqref{eq:starparabolic} is invariant under integer
shifts $\theta(x+1)=\theta(x)+m$. This is no longer true for the
$\IZ_4$ elliptic twist, which requires the full machinery of Morita
equivalence of noncommutative tori to explain the invariance of the
Yang--Mills theory on the D2-brane; this necessitates, in particular, the
non-trivial actions of the elliptic monodromy on the remaining moduli of the
gauge theory described above.

We can give a physical picture for this distinction by including
the background magnetic flux modulus $\Phi$
which twists the vacuum of the noncommutative gauge theory as
discussed in \S\ref{sec:constantB}. It shifts the noncommutative
field strength tensor $F_{\star}$ defining the Yang--Mills action by
a closed two-form $\Phi$ on the D2-brane worldvolume to give
\bea\label{eq:FstarPhiell}
\mathcal{F}=F_{\star} + \Phi \ .
\eea
Under a Morita transformation \eqref{eq:SL2Ztheta} with
monodromy $\cM$, the magnetic flux $\Phi$ transforms as 
\bea
\cM[\Phi] = -(c\,\theta+d)^2\, \Phi + c\, (c\,\theta+d) \ .
\eea
This also affects the spectrum of D-brane charges on the $T^2$ fibres: a generic
configuration $(n,q)\in\IZ^2$ of $n$ D2-branes wrapping $T^2$ with $q$ units of D0-brane charge, realised as a background magnetic charge $q$ in $U(n)$
noncommutative Yang--Mills theory, transforms as a vector under $SL(2,\IZ)_\theta$ to
\bea\label{eq:Moritacharges}
\cM\bigg[\begin{pmatrix} n \\ q \end{pmatrix}\bigg]
= \begin{pmatrix} a & b \\ c & d \end{pmatrix} 
\begin{pmatrix} n \\ q \end{pmatrix} \ .
\eea

For the parabolic monodromies \eqref{eq:parmonodromy} this has no effect; in that case $\Phi(x+1)=\Phi(x)$
and the single D2-brane charge with $(n,q)=(1,0)$ that we have
considered is mapped to itself. Thus we can consistently set $\Phi=0$ for the parabolic
 case and simply write the standard noncommutative Yang--Mills
action in terms of $F_{\star}$ and single-valued gauge fields on the
$T^2$ fibre. 

In contrast, for the $\IZ_4$ elliptic
monodromy \eqref{eq:Z4monodromy} the magnetic flux $\Phi$ on the D2-brane transforms as
\bea
\Phi(x+1) = -\theta(x)^2\, \Phi(x) + \theta(x) \ ,
\eea
while the single D2-brane charge $(1,0)$ that we started with is mapped to a single
D0-brane charge $(0,-1)$; the notation $(n,q)$, as above, denotes $n$ units of D2-brane charge and $q$ units of D0-brane charge. This is solved by
\bea\label{eq:Phielliptic}
\Phi = -\frac{x}{\theta(x)} \ \dd y^1\wedge \dd y^2\ ,
\eea
which should be properly incorporated in \eqref{eq:FstarPhiell} in
order to obtain the correct compactification of the open string
non-geometric background via shifts in $x$. Choosing the real part of
the fixed modular parameter $\tau^\circ$ to be $\tau_1^\circ=-1$, the
two-form \eqref{eq:Phielliptic} then correctly interpolates
between the D0-brane charges $\Phi(0)=0 $ and $\Phi(1)=-1$, implying that single-valued gauge fields are mapped to multi-valued gauge fields on the D2-brane under the monodromy in the base coordinate $x\in S^1$~\cite{Ambjorn2000}.

This effect may be interpreted in terms of open string boundary
conditions. By observing that the bivector $\theta(x)$ is constant on
the D2-brane worldvolume, we can easily adapt the description of open
string ground states given in~\cite{Seiberg1999} as Morita equivalence
bimodules for a noncommutative torus, as is also done by~\cite{Lowe2003}
in a different context. We start with a single D2-brane placed at
some fixed point $x\in S^1$, wrapping the $T^2$ fibre. Since open
strings interact by concatenation of paths, the space of open string ground
states on the D2-brane forms an algebra~$\cA_x$ descending from the
algebra of open string tachyon vertex operators in the decoupling
limit, which is precisely the algebra of functions on the
noncommutative torus $T^2_{\theta(x)}$ with the star-product
\eqref{eq:starparabolic}. 

Consider now a string winding once around the $x$-circle, viewed in the covering space $\IR$ over $S^1$ as an open string
stretching with its left end on the $(1,0)$ D2-brane at the point $x$ and with its right end
on some other configuration $(n,q)$ of D2--D0-branes placed at the
point $x+1$, with the $n$ D2-branes wrapping $T^2$
and the $q$ D0-branes located at $y^1=y^2=0$ on $T^2$. The configuration $(n,q)$ comes with its own algebra $\cA_{x+1}$ of open string ground
states, identified as the algebra of functions on $T^2_{\theta(x+1)}$ for a suitable monodromy \eqref{eq:SL2Ztheta} of the noncommutativity parameter determined by the transformation \eqref{eq:Moritacharges} of D-brane charges. Quantisation of the stretched open string with these boundary
conditions in the decoupling limit gives a space of states $\cH_{n,q}$
which is a left module for the algebra $\cA_x$, acting on the left end
of the open string, and a right module for $\cA_{x+1}$, acting on the
right end of the open string. The actions of $\cA_x$ and $\cA_{x+1}$
commute because they act at opposite ends of the open string, and
together they generate the complete algebra of observables on the open
string tachyon ground states, acting irreducibly on $\cH_{n,q}$. This implies that
the algebra $\cA_{x+1}$ is the commutant of $\cA_x$ in this space (and
vice-versa), the maximal algebra of all operators on $\cH_{n,q}$ that
commute with $\cA_x$. In~\cite{Seiberg1999} it is shown that the space
$\cH_{n,q}$ thus defines a Morita equivalence bimodule over $\cA_x\times\cA_{x+1}$
in this sense, which mathematically implements the Morita
equivalence between noncommutative tori whose algebras of functions
are $\cA_x$ and $\cA_{x+1}$; roughly speaking, this implies that there
is a bijective mapping between the ``gauge bundles'' over the
noncommutative tori $T^2_{\theta(x)}$ and~$T^2_{\theta(x+1)}$.

The parabolic Morita duality above dictates that the right end of the
open string should also land on a single $(1,0)$ D2-brane at $x+1$. In this
case the Morita equivalence is trivial: The space of open string
ground states $\cH_{1,0}$ in the decoupling limit is simply a copy of
the algebra $\cA_x$ itself, or the \emph{free} bimodule over
$\cA_x$, with the algebra $\cA_x$ of functions on the noncommutative torus
$T^2_{\theta(x)}$ acting from the left and the algebra
$\cA_{x+1}=\cA_x$ of functions on 
$T^2_{\theta(x+1)}=T^2_{\theta(x)}$ acting from the right, both via
the star-product \eqref{eq:starparabolic}; this identifies $\cH_{1,0}$
as the space of functions on an ordinary torus $T^2$. 
Thus the
parabolic monodromies recover the standard free 
bimodule over $\cA_x$.

In contrast, the $\IZ_4$ elliptic Morita duality dictates that the right end
of the open string should land on a single $(0,-1)$ D0-brane at
$x+1$. The space of open string
ground states $\cH_{0,-1}$ in this case comes from quantising a
two-dimensional phase space, which is the cover of the torus $T^2$ with the
Poisson bracket $\{y^1,y^2\}=\theta(x)$. 
This may be identified as an
algebra of functions on $\IR$ in a Schr\"odinger
polarisation in which the algebra $\cA_x$ 
acts
on $\cH_{0,-1}$, regarded as the space of functions of $y^2$, by representing $y^2$ as multiplication by $y^2$ and $y^1$ as the
derivative $\ii\theta(x)\, \frac\partial{\partial y^2}$. The commutant
of $\cA_x$ in $\cH_{0,-1}$ is generated by operators given as
multiplication by $y^2/\theta(x)$ together with the derivative
$\ii\frac\partial{\partial y^2}$, which quantise the Poisson bracket
$\{y^1,y^2\}=\theta(x)^{-1}$. This gives the standard Morita
equivalence bimodule over $\cA_x\times\cA_{x+1}$~\cite{Seiberg1999}, with the algebra
$\cA_x$ of functions on the
noncommutative torus $T^2_{\theta(x)}$ acting from the left and the
algebra $\cA_{x+1}$ of functions on 
$T^2_{\theta(x+1)}=T^2_{-1/\theta(x)}$ acting from the right.

More general Morita
transformations of the D-brane charges, taking the initial
configuration of charges $(1,0)$ to a
configuration $(1,1)$ with a unit of D0-brane charge inside a
D2-brane, are possible with the $\IZ_3$ elliptic monodromy, and can
be similarly interpreted on the space $\cH_{1,1}$ of sections of a
line bundle over $T^2$ with Chern number $1$~\cite{Seiberg1999}.

\subsection{D2-brane theory at the orbifold point}

In twisted dimensional reductions, the scaling limits discussed in
this section describe D-branes with $x$-dependent noncommutativity
parameters $\theta(x)$ coupled to gauged supergravity. One of the most interesting features of elliptic twists, as compared
to parabolic twists, 
{is that they each admit  a fixed point in 
moduli space at which the 
twisted reduction reduces to an orbifold reduction and so gives an exact string theory realisation~\cite{Dabholkar2002}.  The fixed point for a given elliptic twist is at a   minimum of the corresponding Scherk--Schwarz potential at which the potential vanishes, and so gives a  stable compactification to Minkowski space~\cite{Dabholkar2002}. The twist $\gamma (x)$ at the fixed point is independent of
$x$ and the monodromy $\cM$ generates a cyclic group of order $p$ for some integer $p$, $\cM^p=1$. The twisted reduction at the fixed point then is realised as a
$\IZ _p$ orbifold of the theory compactified on $T^3$. This is given by the 
 compactification on $S^1\times T^2$ orbifolded by the action of  $\cM$ on the $T^2 $ conformal field theory
together with a shift 
$x\mapsto x+\frac1p$
 of the coordinate $x$ of the $S^1$.} 
   In
particular, from \eqref{eq:elliptictaux} it follows that
$\tau^\circ=\ii$ is a fixed point of the $SL(2,\IZ)$ transformation
generated by \eqref{eq:ellipticmonodromy} for $U=\unit$: In that case
$\tau(x)=\ii$ independently of $x\in S^1$, and the minimum of the potential gives a Minkowski vacuum. {The construction is a $\IZ_4$ orbifold of the compactification
on $S^1\times T^2_{\tau(x)=\ii}$ with the $\IZ_4$ twist of the conformal field theory on the $T^2$ with $\tau^\circ=\ii$ accompanied by a shift $x\mapsto x+\frac14$.}

At this point it is not
possible to decouple open and closed string modes on the D2-brane,
which would require a scaling limit $\tau_2^\circ\to0$. In fact, at
this point the $B$-field vanishes and the closed string metric
\bea
g_{\tau(x)=\ii} = (2\pi\,r \, \dd x)^2 + \big(2\pi\,r_1 \, \dd y^1\big)^2
+ \big(2\pi\,r_2 \, \dd y^2\big)^2 
\eea
coincides exactly with the $T^3$ metric for finite string
slope $\alpha'$ and area $A$. As expected from the open-closed
transformation \eqref{eq:openstringGTheta} with $B=0$, in this case
the open and closed string metrics on the D2-brane coincide while the
noncommutativity vanishes, $\Theta=0$. The dilaton is given by
$\e^{\phi_{\tau(x)=\ii}} = (2\pi\,\alpha'/A)^{1/2}$ and hence the Yang--Mills coupling at
the fixed point is the $\alpha'$-independent constant
\bea
\big(g^{\phantom{\dag}}_{\tiny\rm{YM}}\big)_{\tau(x)=\ii} = \bigg(\frac{2\pi \,
  g_s^2}A\bigg)^{1/4} \ .
\eea
Thus in this case the closed string geometry is identical to
the open string geometry, and the worldvolume gauge theory on the D2-brane is that
of an ordinary \emph{commutative} supersymmetric Yang--Mills theory on a flat
torus $T^2\subset T^3$. The same is expected to be true for the
$\IZ_3$ (and $\IZ_6$) twist at the fixed point
$\tau^\circ=\e^{\pi\ii/3}$, which can {be viewed as a $\IZ_3$ orbifold and as a toroidal
reduction with magnetic flux.}

\section{D-branes and doubled twisted torus geometry\label{sec:doubled}}

Having understood the non-geometric T-fold backgrounds, our
aim now is to study D-branes in the essentially doubled  space
 obtained by T-duality in the $x$-direction. However, T-duality along the vector field $\frac\partial{\partial x}$ is problematic because the background depends explicitly on $x\in S^1$: The vector field $\frac\partial{\partial x}$ does not generate an isometry of the torus bundle and the Buscher construction can no longer be applied. 
For such cases, we use a generalised T-duality~\cite{Dabholkar2005} which takes a background with dependence on $x$ to an essentially doubled background in which the fields depend on the coordinate $\wx$ of the T-dual circle and so are problematic to interpret in conventional terms.

The reduction with duality twist by an $O(d,d;\IZ)$ monodromy around the $x$-circle is generalised to a
twisted construction with both a twist along $x$ and along its dual coordinate $\wx$, so that the dependence of the moduli $E(x,\wx)$ is through a local section $\gamma:S^1\times S^1\to O(d,d)$ given by
\bea
\gamma(x,\wx) = \exp(x\,M)\, \exp(\wx\,\wM\,) \ ,
\eea
with commuting mass matrices $M,\wM$ and with the corresponding monodromies
\bea
\cM = \exp(M) \qquad \mbox{and} \qquad \wcM = \exp(\wM\,)
\eea
both valued in $O(d,d;\IZ)$. A general non-geometric reduction then
gives rise to a torus bundle with doubled fibres $T^{2d}$, and
coordinates $y^a,\wy_a$, $a=1,\dots,d$, over a doubled base $S^1\times S^1$,
with coordinates $x,\wx$, such that a generalised T-duality along the
vector field $\frac\partial{\partial x}$ takes a $T^{2d}$-bundle over the $x$-circle to
a $T^{2d}$-bundle over the dual $\wx$-circle. In the remainder of this paper we describe the noncommutative Yang--Mills theories on D-branes
in a doubled geometry of this type which is the natural lift of the
twisted torus backgrounds considered in \S\ref{sec:twistedtorus}. 

We will formulate the theory using a doubled geometry with coordinates $(x,\wx, y^a,\wy_a)$.
Such a doubled formulation was first proposed in~\cite{Hull:2007jy} and developed in~\cite{DallAgata:2007egc,Hull2009,ReidEdwards2009}, replacing the doubled torus with a twisted version, the doubled twisted torus.
This $2n$-dimensional doubled geometry incorporates all the dual forms
of the $n$-dimensional background, with the different backgrounds
arising {from different polarisations, which give different
  $n$-dimensional \lq slices'.} This 
doubled geometry has been recently discussed in~\cite{Marotta2018}.

\subsection{The doubled twisted torus}

Following~\cite{Hull2009,ReidEdwards2009}, we extend the twisted torus
$X=G_\IZ\setminus G_\IR$ of \S\ref{sec:twistedtorus} to the
$2n$-dimensional doubled twisted torus
\bea
\ccX = \ccG_\IZ\setminus \ccG_\IR
\eea
where the $2n$-dimensional non-compact Lie group $\ccG_\IR$ is the
cotangent bundle $\ccG_\IR = T^*G_\IR = G_\IR\ltimes\IR^n$;
 this is a
Drinfeld double, and $\ccG_\IZ$ is a discrete cocompact subgroup of
$\ccG_\IR$. The local structure of $\ccX$ is given by the Lie algebra
of $\ccG_\IR$ whose generators $\cJ_M$, $M=1,\dots,2n$, have brackets
\bea
[\cJ_M,\cJ_N] = t_{MN}{}^P\, \cJ_P \ , 
\eea
with structure constants $t_{MN}{}^P$ satisfying the Jacobi identity
$t_{[MN}{}^Q\, t_{P]Q}{}^T=0$. The Lie algebra admits an
$O(n,n)$-invariant constant symmetric bilinear form $\eta_{MN}$ of
signature $(n,n)$, and so  $\ccG_\IR$ is a $2n$-dimensional subgroup of $O(n,n)$.
 The generators $\cJ_M$ consist of $J_m=\{J_x,J_a\}$,
$a=1,\dots,d$, $m=1,\dots,n$, generating the   $G_\IR$ subgroup, and $\wJ^m=\{\wJ^x,\wJ^a\}$
generating the $\IR^n$ subgroup.
Then 
\bea
\cJ_M = 
\begin{pmatrix}
J_m \\ \wJ^m
\end{pmatrix}
\eea
is formally an $O(n,n)$-vector and in this basis the $O(n,n)$-invariant metric is
\bea
\eta = 
\begin{pmatrix}
0 & \unit \\
\unit & 0
\end{pmatrix} \ .
\eea
The Lie algebra is
\bea
[J_a,J_x] = M_a{}^b \, J_b \ , \qquad [J_a,J_b]=0 \quad & \mbox{and} & \quad
[\wJ^a,\wJ^x]=0=[\wJ^a,\wJ^{\,b}] \ , \nonumber \\[4pt]
[J_a,\wJ^{\,b}]=-M_a{}^b \, \wJ^x \ , \qquad [J_x,\wJ^a] = M_b{}^a \,
\wJ^{\,b} \quad & \mbox{and} & \quad [J_a,\wJ^x] = 0 = [J_x,\wJ^x] \ .
\eea
This has the Drinfeld double form
\bea
[J_m,J_n]=f_{mn}{}^p\,J_p \ , \quad   [\wJ^m,\wJ^n]=0 \qquad \mbox{and} \qquad
[J_n,\wJ^m]=-f_{np}{}^m\,\wJ^p
\label{eq:falg}
\eea
where 
$f_{mn}{}^p$ are the structure constants for  $G_\IR$.

The group manifold of $\ccG_\IR$ is parameterised by coordinates
$(x,y^1,\dots,y^{d})$ on $G_\IR$ as in \eqref{eq:GR}, and coordinates
$(\wx,\wy_1,\dots,\wy_{d})$ on $\wG_\IR=\IR^n$. The quotient by the action of
the discrete cocompact subgroup
$\ccG_\IZ$ of the $2n$-dimensional group $\ccG_\IR$ 
results in the  compact space $\ccX$.
The T-dual coordinates $(\wx,\wy_a)$
are all periodic and so parameterise an $n$-torus $T^n$, so that
$\ccX$ admits a $T^n$ fibration with fibre coordinates $(\wx,\wy_a)$
as well as the $T^{2d}$ doubled torus fibration with fibre coordinates $(y^a, \wy_a)$.

The action of $\ccG_\IZ$ induces a
monodromy in $O(d,d;\IZ)\subset GL(2d,\IZ)$ acting geometrically by a
large diffeomorphism of the doubled torus fibres as
\bea
\begin{pmatrix} y^a \\ \wy_a \end{pmatrix}
\xrightarrow{\ x\mapsto x+1 \ } \begin{pmatrix} \cM^{-1} & 0 \\
0 & \cM^\top \end{pmatrix} \begin{pmatrix} y^a \\ \wy_a \end{pmatrix}
\ ,
\eea
together with the $GL(d,\IZ)$ monodromy
\bea
{\small\begin{pmatrix} \wx \\ \wy_1 \\ \vdots \\ \wy_{d-1}\end{pmatrix}}
\xrightarrow{\ y^{d}\mapsto y^{d}+1 \ } \cM^{-1} 
{\small\begin{pmatrix} \wx \\ \wy_1 \\ \vdots \\ \wy_{d-1}\end{pmatrix}}
\eea
acting geometrically as a large diffeomorphism on the T-dual
torus. Here $\cM$ is the monodromy matrix of the twisted torus $X$ from
\S\ref{sec:twistedtorus}, given explicitly for $n=3$ ($d=2$) by \eqref{eq:parmonodromy} in the
case of parabolic twists and by \eqref{eq:ellipticmonodromy} for
elliptic twists.

The Maurer--Cartan one-forms \eqref{eq:MC1forms} lift to
left-invariant forms on $\ccG_\IR$, but  $G_\IR$ acts
non-trivially on $\wG_\IR$ so one needs to ``twist'' the
left-invariant one-forms $\dd\wx,\dd\wy_a$ of $\wG_\IR$ when lifting
them to $\ccG_\IR$. A basis of left-invariant one-forms on $\ccG_\IR$
is then given by
\bea
\zeta^x = \dd x  \quad & \mbox{and} & \quad \zeta^a = \gamma(x)^a{}_b \,
\dd y^b  \ , \nonumber \\[4pt]
\weta_x = \dd\wx - M_a{}^b \, \wy_b \, \zeta^a \quad & \mbox{and} &
\quad \weta_a = \dd\wy_a + M_a{}^b \, \wy_b \, \zeta^x \ .
\eea
The action of $\ccG_\IZ$ is compatible with $\wG_\IR$, so that the
quotient $\wG_\IR\setminus\ccX$ is well-defined and corresponds to the
$n$-dimensional twisted torus $X=G_\IZ\setminus G_\IR$. In this way
the conventional spacetime description is obtained for the natural
polarisation associated to the coset $\wG_\IR\setminus\ccG_\IR$, which
corresponds to the natural projection on the cotangent bundle $T^*X=X\times\IR^n$.

We write coordinates on the quotient $\ccX$ as
$\bX^I=(x,y^a,\wx,\wy_a)$ with $a=1,\dots,d$ and
$I=1,\dots,2n$, and the one-forms $\zeta^m=\{\zeta^x,\zeta^a\}$ and $\weta_m=\{\weta_x,\weta_a\}$ with $m=1,\dots,n$
collectively as
\bea
\cP^M = \cP^M{}_I \, \dd\bX^I \ .
\eea
We will sometimes denote these coordinates as $\bX^M=(x^m,\wx_m)$,
and write a general group element $g\in\ccG_\IR$ as
\bea
g = \widetilde{h}\, h
\eea
where 
\bea
h=\exp(x^m\, J_m) \qquad \mbox{and} \qquad \widetilde{h}=\exp(\wx_m\,
\wJ^m) \ .
\eea
{The   left-invariant one-forms $\cP^M$, $M=1,\dots,2n$ 
satisfy the
Maurer--Cartan
equations
\bea
\dd\cP^M + \mbox{$\frac12$} \, t_{NP}{}^M \, \cP^N\wedge\cP^P = 0 \ ,
\eea
so that the $2n$-manifold $\ccX$ is parallelisable.}
We further
introduce a constant  (independent of the coordinates $\bX$ on $\ccX$) metric given by a $2n\times2n$ symmetric matrix $\sfM_{MN}$
satisfying
\bea
\sfM_{MN} = \eta_{MP} \, (\sfM^{-1})^{PQ} \, \eta_{QN} \ ,
\eea
so that it
 parameterises
the left coset $\big(O(n)\times O(n)\big)\setminus O(n,n)$.
The matrix $\sfM_{MN}$ constitutes the moduli of the doubled twisted torus, and becomes a matrix of scalar fields on dimensional reduction, giving a non-linear sigma-model with  target space $\big(O(n)\times O(n)\big)\setminus O(n,n)$.

A natural  left-invariant metric and three-form on $\ccX$ are then defined by
\bea
\dd s^2_\ccX = \sfM_{MN} \, \cP^M \, \cP^N = \cH_{IJ} \, \dd\bX^I \,
\dd\bX^J
\eea
and
\bea
\cK = \mbox{$\frac16$} \, t_{MNP} \,
\cP^M\wedge\cP^N\wedge\cP^P = \mbox{$\frac16$} \, T_{IJK} \,
\dd\bX^I\wedge\dd\bX^J\wedge\dd\bX^K \ ,
\eea
where the doubled metric $\cH_{IJ}:=\sfM_{MN} \, \cP^M{}_I \, \cP^N{}_J$ obeys $\cH_{IJ} = \eta_{IK} \, (\cH^{-1})^{KL} \, \eta_{LJ}$, with
    $\eta_{IJ} = \eta_{MN} \, \cP^M{}_I \, \cP^N{}_J$, while $T_{IJK}:=
    t_{MNP} \, \cP^M{}_I \, \cP^N{}_J \, \cP^P{}_K$ with $t_{MNP}:=
    \eta_{MQ} \, t_{NP}{}^Q$  totally antisymmetric. The
    Wess--Zumino three-form $\cK$ is closed, $\dd\cK=0$, by virtue of
    the Jacobi identity $t_{[MN}{}^Q \, t_{P]Q}{}^T=0$.

The natural action of $O(n,n)$ on  the tangent bundle of $\ccG_\IR$ then gives
\bea
 \dd \bX^I &\longmapsto& (\cO^{-1})^I{}_J \,  \dd \bX^J \ , \nonumber \\[4pt]
\cP^M &\longmapsto& (\cO^{-1})^M{}_N \, \cP^N \ , \nonumber \\[4pt]
\cJ_M &\longmapsto&   \cJ_N \, \cO^N{}_ M \ ,
\label{eq:O33transf}\eea
for $\cO\in O(n,n)$. 
This is  essentially the $O(n,n)$ structure group of generalised
geometry, acting on the generalised tangent bundle $TX\oplus T^*X$ of
the $n$-dimensional twisted torus $X$.
Note, however, that $O(n,n;\IZ)$ is {\it not} a symmetry in this case. 
Consider the subgroup $GL(n,\IZ) \subset O(n,n;\IZ)$.
For $T^n$, the group of large diffeomorphisms is $GL(n,\IZ)$, but for
the $n$-dimensional twisted torus $X$, $GL(n,\IZ)$ is not a symmetry,
although the subgroup $GL(d,\IZ)$ acting on the $T^{d}$ fibres is. For
string theory on $X$, or any of its T-duals, there is an $O(d,d;\IZ)$
T-duality symmetry acting in the conformal field theory on the
$T^{d}$ fibres; in the doubled formalism, this acts as a
diffeomorphism  on the $T^{2d}$ doubled torus fibres of $\ccX$. For
$n=3$ ($d=2$), this $O(2,2;\IZ)$ acting on $(y^a,\wy_a)$ recovers the T-duality
    orbits of the three-dimensional twisted torus from \S\ref{sec:Tfoldparabolic} and \S\ref{sec:Tfoldelliptic}. 
For $\cO \in  O(d,d;\IZ) \subset O(n,n)$, the action \eqref{eq:O33transf} extends to
\bea
  \bX^I &\longmapsto& (\cO^{-1})^I{}_J \,   \bX^J \ , 
 \nonumber \\[4pt]
\cH_{IJ}(\bX) &\longmapsto& \cO_I{}^K \, \cH_{KL}(\cO^{-1} \bX) \, \cO^L{}_J \ , \nonumber \\[4pt]
T_{IJK}(\bX) &\longmapsto& T_{LMN}(\cO^{-1} \bX) \, \cO^L{}_I \,
\cO^M{}_J \, \cO^N{}_K \ .  
\label{eq:O33transfa}
\eea

In~\cite{Hull2009}, it was proposed that generalised T-duality acts in the same way, for certain  other $\cO \in    O(n,n)$.
In particular, it was proposed that the generalised T-duality  in the $x$-direction is given by an $O(n,n)$-transformation $\cO_x$, which for $n=3$ reads as
\bea
{\cal O}_x
=
{\small\begin{pmatrix}
  0 & 0 & 0 & 1 & 0 & 0 \\
  0 & 1 & 0 & 0 & 0 & 0 \\
  0 & 0 & 1 & 0 & 0 & 0 \\
  1 & 0 & 0 & 0 & 0 & 0 \\
  0 & 0 & 0 & 0 & 1 & 0 \\
  0 & 0 & 0 & 0 & 0 & 1 
\label{eq:Ox}\end{pmatrix}}
\eea
acting on 
\bea
  \mathbb{X}^{{I}}
  =
{\small\begin{pmatrix}
x\\  y^1 \\ y^2 \\ \wx \\ \wy_1 \\ \wy_2 
\end{pmatrix}}
\eea
corresponding to the exchange $x\leftrightarrow \wx$.

The transformation
$\cJ_M \longmapsto   \cJ_N \, \cO^N{}_ M$
changes the split of the Lie algebra generators $\cJ_M$ into $J_n,\wJ^m$ and so changes the form of the algebra 
\eqref{eq:falg}
to an algebra of the form 
\bea
[J_m,J_n]&=&f_{mn}{}^p\, J_p+h_{mnp}\, \wJ^p \ , \nonumber \\[4pt]
[\wJ^m,\wJ^n]&=&Q_p{}^{mn}\, \wJ^p+R^{mnp}\, J_p \ , \nonumber\\[4pt]
[J_n,\wJ^m]&=&-f_{np}{}^m\, \wJ^p+ Q_n{}^{mp}\, J_p \ , 
\label{eq:Ralg}
\eea
where the various tensors are determined by the choice of $\cO$, and
are sometimes referred to as fluxes.

In~\cite{Hull2009}, these dualities were interpreted in terms of the choice of  polarisation, generalising the picture in \cite{Hull2004}.
A polarisation splits the tangent bundle $T\ccX$  of $\ccX$ at each point into an $n$-dimensional physical subspace $\mit\Pi$  and an $n$-dimensional dual subspace $\widetilde{\mit\Pi}$, and the issue is whether the 
split of the tangent vectors defines an $n$-dimensional submanifold (at least locally) which can be viewed as a patch of spacetime.
If the spaces $\mit\Pi$ and $\widetilde{\mit\Pi}$ define an integrable distribution on $\ccX$, then there is locally a physical subspace of $\ccX$ and the background is locally geometric.
If the distribution is non-integrable, then there is no such local spacetime and the background is not geometric even locally; it is essentially doubled.
The polarisation splits the $2n$ Lie algebra generators $\cJ_M$ into two sets of $n$ generators,  the $J_m$ and the $\wJ^m$.
The integrability condition is that the  $\wJ^m$ generate a subgroup $\wG_\IR \subset \ccG_\IR$. Then the physical spacetime is defined by the quotient by 
$\wG_\IR$. The covering space for $\ccX$ is  $\ccG_\IR$ and the covering space for the physical subspace is the coset
$\ccG_\IR /\wG_\IR$. If the action of $\ccG_\IZ$ is compatible with the action of $\wG_\IR$, then the background is geometric and given globally by a double quotient of 
$\ccG_\IR$ by $\ccG_\IZ$ and $\wG_\IR$.
If it is not, then the result is a T-fold, with local $n$-dimensional patches given by patches of $\ccG_\IR /\wG_\IR$.
A T-duality transformation is then interpreted as a change of polarisation, changing the physical subspace within the doubled space, and can be realised as the action of the operator $\cO$ on the projectors defining the polarisation~\cite{Hull2009}.

The vielbeins $\cP^M{}_I$
    are maps $\cP:O(n,n)\to O(n)\times O(n)$ and can be brought to
    lower block-triangular form by an $O(n)\times O(n)$ transformation
    to get
\bea
\cP = 
\begin{pmatrix}
e & 0 \\
-(2\pi\,\alpha') \, e^{-1} \, B & (2\pi\,\alpha') \, e^{-1}
\end{pmatrix} \ ,
\eea
where $e$ is the vielbein for the spacetime metric $g= e^\top \,
e$, and $B$ is the NS--NS two-form potential. By
choosing the simple background $\sfM_{MN} = \delta_{MN}$, the
doubled metric can be written as
\bea
\cH = 
\begin{pmatrix}
g-(2\pi\,\alpha')^2 \, B \, g^{-1} \, B & (2\pi\,\alpha')^2 \, B \,
g^{-1} \\
-(2\pi\,\alpha')^2 \, g^{-1} \, B & (2\pi\,\alpha')^2 \, g^{-1}
\end{pmatrix} \ .
\label{eq:doubledmetric}\eea
The expressions for general moduli $\sfM_{MN} $ are given in~\cite{Hull2009}.

If $\widetilde g$ denotes the dual metric arising from an $O(n,n)$
transformation \eqref{eq:O33transfa} of \eqref{eq:doubledmetric}, then
the dilaton transforms as
\bea
\e^\phi\longmapsto \bigg(\frac{\det\widetilde{g}}{\det
  g}\bigg)^{1/4} \, \e^{\phi} \ .
\label{eq:O33dilaton}\eea

\subsection{D-branes in the doubled twisted torus}

D-branes in the
doubled picture were discussed for the  doubled torus and  for doubled torus fibrations in~\cite{Hull2004,Lawrence2006},
and this was extended to the doubled
twisted torus 
in~\cite{Albertsson2008}.
Following~\cite{Albertsson2008},
 let us now describe D-branes in the
doubled twisted torus geometry. 

The starting point is the doubled sigma-model which was introduced
in~\cite{Hull2009} for maps
embedding a closed string worldsheet $\Sigma$ in
$\ccX$. These maps pull back the one-forms $\cP^M$ to one-forms
$\hat\cP^M$ on $\Sigma$. Introducing a three-dimensional manifold $V$
with boundary $\partial V=\Sigma$ and extending the maps to $V$, the
sigma-model is defined by the action
\bea\label{eq:sigmadouble}
S_\ccX = \frac14\, \oint_\Sigma\, \sfM_{MN}\,
\hat\cP^M\wedge\ast\hat\cP^N + \frac1{2}\, \int_V\, \hat\cK \ ,
\eea 
where $\hat\cK$ is the pullback of the Wess--Zumino three-form $\cK$ to $V$ and $\ast$ is the
Hodge duality operator on $\Sigma$. To recover the ordinary non-linear
sigma-model on a physical target space, this doubled sigma-model is subjected to the
self-duality constraint
\bea\label{eq:selfduality}
\hat\cP^M = \eta^{MP}\, \sfM_{PN}\ast\hat\cP^N
\eea
which eliminates half of the $2n$ degrees of freedom by restricting $n$ of
them to be right-moving and $n$ of them to be left-moving on $\Sigma$. In~\cite{Hull2009} this
constraint was imposed by choosing a polarisation and then gauging the
sigma-model. 

In the case that {the structure constants  $R^{mnp}$ in \eqref{eq:Ralg} vanish}, so that
the $\wJ^m$ generate a subgroup $\wG_\IR\subset\ccG_\IR$, then the
reduction to the physical subspace can be achieved by gauging the
action of $\wG_\IR$. On quotienting by the discrete subgroup
$\ccG_\IZ$ and eliminating the worldsheet gauge fields, one obtains a
standard non-linear sigma-model on a target space described locally by the coset
$\ccG_\IR/\wG_\IR$ with coordinates $(x,y^1,\dots,y^{d})$, with metric and $B$-field given from the
generalised metric \eqref{eq:doubledmetric}, and with physical $H$-field
strength given by
\bea
H=\dd B \ .
\eea
For more general doubled groups that are not Drinfeld doubles, the equation for $H$ has further terms which are given in~\cite{Hull2009}. 

On the other hand, if {the structure constants  $R^{mnp}$ in \eqref{eq:Ralg} are} non-zero, then the sigma-model will depend explicitly on both $x$ and
$\wx$. In this essentially doubled case, the metric and $H$-field
strength depend on both $x$ and
$\wx$, and it appears that there is no interpretation of the sigma-model in terms of a
conventional $n$-dimensional spacetime.

In~\cite{Albertsson2008} it was shown that the same sigma-model action
\eqref{eq:sigmadouble} can be used to describe the embedding of an
open string worldsheet $\Sigma$ in the doubled space $\ccX$. In this
case one must specify boundary conditions by demanding that the string maps
should send the boundary $\partial\Sigma$ of $\Sigma$ to a given
submanifold $\mathscr{W}\subset\ccX$, the worldvolume of a D-brane
in the doubled space $\ccX$. This requires that the embedding of the
boundary $\partial V$ of the three-dimensional manifold $V$ is the sum
of the embedding of $\Sigma$ with some chain
$\mathscr{D}\subseteq\mathscr{W}$, and consistency of the Wess--Zumino
term in \eqref{eq:sigmadouble} requires that the pullback of the
three-form $\cK$ to $\mathscr{D}$ vanishes, $\cK|_{\mathscr{D}}=0$. One
can then analyse the boundary equations of motion as well as the
self-duality constraint \eqref{eq:selfduality} with these
conditions. In~\cite{Albertsson2008} it is shown that as a result the
worldvolume $\mathscr{W}$ of a D-brane in the doubled
twisted torus is a subspace of $\ccX$ which is
maximally isotropic with respect to the $O(n,n)$-invariant metric $\eta$. Choosing a
polarisation then picks out physical worldvolume
coordinates, so that the physical D-brane wraps that part of the physical space which intersects the
generalised D-brane subspace $\mathscr{W}$. D-branes in the doubled space are specified by
complementary Dirichlet and Neumann
projectors that respectively project the tangent bundle of  $\ccX$ at
each point into
subspaces
 normal and
tangential to the worldvolume wrapped by the D-brane. Both subspaces
are null with respect to $\eta$, and they are mutually orthogonal to each
other with respect to the doubled metric $\cH$. The Neumann projector
moreover satisfies an integrability condition ensuring that the
D-brane worldvolume $\mathscr{W}$ is locally a smooth submanifold of
$\ccX$. The vanishing of the Wess--Zumino three-form $\cK$ on $\mathscr{W}$
further constrains the structure constants $t_{MN}{}^P$
of the Lie algebra of $\ccG_\IR$ which restricts the orientation of
the D-brane in~$\ccX$. 

This construction implies, in particular, that for each Neumann
condition there is a corresponding Dirichlet condition. Thus there are always $n$ Neumann directions and
$n$ Dirichlet directions on the doubled twisted torus $\ccX$, and
these directions each form a null subspace of $\ccX$. As a consequence, any D-brane in a physical $n$-dimensional polarisation 
always arises from a D$n$-brane in the extended $2n$-dimensional doubled
geometry. 

As before, we shall study the cases with $n=3$ in detail. Starting from the three-dimensional spacetime polarisation
above onto the twisted torus $X$, we will follow the T-duality orbits
of D-branes in $\ccX$. 
The D-brane projectors transform under the action of the T-duality operator $\cO$, and the
possible D-branes in the various T-duality frames are classified using the doubled twisted torus formalism by~\cite{Albertsson2008}. In particular,
some anticipated worldsheet classification results are confirmed explicitly in
this way; for example, it is known that D3-branes cannot wrap the
three-torus $T^3$
with non-zero $H$-flux due to the Freed--Witten anomaly~\cite{Freed1999} (because $T^3$
is a spin$^c$-manifold and so anomaly cancellation requires
$m=[H]=W_3(T^3)=0$). 

\subsection{D2-branes on T-folds}

We start by  rederiving the results of \S\ref{sec:Tfoldparabolic} and \S\ref{sec:Tfoldelliptic}
in the doubled picture, which involve T-duality transformations in the $y^a$ direction  corresponding to
$\cO_{y^a}\in O(2,2;\IZ)$. Starting from the twisted torus with metric
\eqref{eq:dsX2general} and vanishing $B$-field, we write the
corresponding doubled
metric from~\eqref{eq:doubledmetric}:
\bea
\cH_f = \begin{pmatrix}
(2\pi\,r)^2 & 0 & 0& 0 & 0& 0 \\
0 & \frac A{\tau_2(x)} & \frac{A\,\tau_1(x)}{\tau_2(x)} & 0 & 0 & 0 \\ 
0 & \frac{A\,\tau_1(x)}{\tau_2(x)} & \frac{A\,|\tau(x)|^2}{\tau_2(x)}
& 0 & 0 & 0 \\
0 & 0 & 0 & \big(\frac{\alpha'}r \big)^2 & 0 & 0 \\
0 & 0 & 0 & 0 & \frac{(2\pi\,\alpha')^2\, |\tau(x)|^2}{A\,\tau_2(x)} & 
-\frac{(2\pi\,\alpha')^2\, \tau_1(x)}{A\,\tau_2(x)} \\
0 & 0 & 0 & 0 & -\frac{(2\pi\,\alpha')^2\, \tau_1(x)}{A\,\tau_2(x)} &
\frac{(2\pi\,\alpha')^2}{A\,\tau_2(x)}
\end{pmatrix} \ ,
\label{eq:cHf}\eea
where the complex structure modulus $\tau(x) = \tau_1(x) + \ii
\tau_2(x)$ is given by \eqref{eq:parabolictaux} in the case of
parabolic twists and by \eqref{eq:elliptictaux} (with $m\in4\,\IZ+1$ and $\vartheta=\frac\pi2$) for the $\IZ_4$
elliptic twist. The Wess--Zumino three-form is given by
\bea
\cK_f = -\tfrac12\, M_a{}^b\, \dd x\wedge\dd\wy_b\wedge\dd y^a \ ,
\eea
where the components of the mass matrix $M$ can be read off from
\eqref{eq:parmonodromy} for the case of parabolic twists and by
\eqref{eq:ellipticmonodromy} (with $U=\unit$, $m\in4\,\IZ+1$ and $\vartheta=\frac\pi2$) for the $\IZ_4$
elliptic twist. In this polarisation one thus finds $H=0$, as expected.
As the only non-vanishing structure constants in
this case are $f_{ax}{}^b=M_a{}^b$, we can wrap a
D1-brane around the torsion one-cycle $\xi_1$ in the doubled geometry~\cite{Albertsson2008},
as previously, and follow its orbits under
T-duality, which are
summarised in {Table}~\ref{tab:orbits}.

\begin{center}
\begin{table}[t]
\begin{center}
	\begin{tabular}{|c|c|c|c|c|}
		\hline {Background} & {Flux}& {D$p$-brane} & {$ \ x \ \ \ 
                                                     y^1 \ \ \, y^2$} &
                                                                     {$
                                                                       \
                                                                       \,
                                                                       \wx
                                                                     \
                                                                     \
                                                                       \ 
                                                                     \wy_{1}
                                                                     \
                                                                     \ 
                                                                     \wy_{2}$}
		\\[4pt] \hhline{|=|=|=|=|=|}
	$T^3$ with $H$-flux &	$H$-flux & D0-brane & $ - \ \ - \ \ - $ & $\times \ \ \times \ \ \times$
		\\[4pt] \hline 
	Nilfold &	$f$-flux & D1-brane & $ - \ \ \times \ \ - $ & $\times \ \ - \ \ \times$
		\\[4pt] \hline 
	T-fold	 &$Q$-flux & D2-brane & $ - \ \ \times \ \ \times$ & $
                                                                 \times
                                                                 \ \ -
                                                                 \ \ -
                                                                 $
\\[4pt] \hline
Essentially doubled	&	$R$-flux & D3-brane & $ \times \ \ \times \ \ \times$ & $
                                                                     -
                                                                     \
                                                                     \
                                                                     -
                                                                     \
                                                                     \
                                                                     - $
\\\hline
	\end{tabular}
\end{center}
\caption{\small The D$p$-brane configurations considered in the
  various T-duality frames of the
  doubled twisted torus geometry for the case of parabolic monodromies. A
  dash denotes a normal (Dirichlet) direction to the D-brane, while a
  cross denotes a worldvolume (Neumann) direction. The number of
  Dirichlet and Neumann directions in each case are both equal to three.}
\label{tab:orbits}\end{table}
\end{center}

To dualise along the vector field $\frac\partial{\partial y^1}$ of $\ccX$, we apply
\eqref{eq:O33transfa} to \eqref{eq:cHf} with
\bea
\cO_{y^1} = {\small\begin{pmatrix}
1 & 0 & 0 & 0 & 0 & 0 \\
0 & 0 & 0 & 0 & 1 & 0 \\
0 & 0 & 1 & 0 & 0 & 0 \\
0 & 0 & 0 & 1 & 0 & 0 \\
0 & 1 & 0 & 0 & 0 & 0 \\
0 & 0 & 0 & 0 & 0 & 1
\end{pmatrix}} \ ,
\eea
which interchanges $y^1$ with $\wy_{1}$ in the doubled coordinates $\bX$
and all fields, leaving all other components invariant. The
transformed doubled metric is given by
\bea
\cH_H = \cO_{y^1}^\top \, \cH_f \, \cO_{y^1} = 
\begin{pmatrix}
(2\pi\,r)^2 & 0 & 0 & 0 & 0 & 0 \\
0 & \frac{(2\pi\,\alpha')^2\,|\tau(x)|^2}{A\,\tau_2(x)} & 0 & 0 & 0 & 
-\frac{(2\pi\,\alpha')^2\,\tau_1(x)}{A\,\tau_2(x)} \\
0 & 0 & \frac{A\,|\tau(x)|^2}{\tau_2(x)} & 0 &
\frac{A\,\tau_1(x)}{\tau_2(x)} & 0 \\
0 & 0 & 0 & \big(\frac{\alpha'}r\big)^2 & 0 & 0 \\
0 & 0 & \frac{A\,\tau_1(x)}{\tau_2(x)} & 0 & \frac A{\tau_2(x)} & 0 \\
0 & -\frac{(2\pi\,\alpha')^2\,\tau_1(x)}{A\,\tau_2(x)} & 0 & 0 & 0 &
\frac{(2\pi\,\alpha')^2}{A\,\tau_2(x)} 
\end{pmatrix} \ .
\eea
Comparing with \eqref{eq:doubledmetric}, we can read off the closed
string metric and $B$-field. For the parabolic monodromy
\eqref{eq:parabolictaux}, 
these agree with \eqref{eq:T3gB} for the geometric
three-torus $T^3$ with constant $H$-flux~\eqref{eq:HT3}, and the Wess--Zumino three-form is given by
\bea\label{eq:KH}
\cK_H=-\mbox{$\frac12$}\, m\, \dd x\wedge\dd y^1\wedge \dd y^2 \ .
\eea 
In this polarisation the generators
$\wJ^m$ generate a maximally isotropic subgroup $\wG_\IR\subset\ccG_\IR$ which
is compatible with the action of $\ccG_\IZ$, so that the quotient
$\wG_\IR\setminus\ccX$ is well-defined and provides a global
description of the three-dimensional compactification
geometry. Altogether we recover the standard non-linear sigma-model
with target space $T^3$ threaded by a constant $H$-flux.

On the other hand, dualising along the vector field $\frac\partial{\partial y^2}$ implements
\eqref{eq:O33transfa} on \eqref{eq:cHf} with
\bea
\cO_{y^2} = {\small\begin{pmatrix}
1 & 0 & 0 & 0 & 0 & 0 \\
0 & 1 & 0 & 0 & 0 & 0 \\
0 & 0 & 0 & 0 & 0 & 1 \\
0 & 0 & 0 & 1 & 0 & 0 \\
0 & 0 & 0 & 0 & 1 & 0 \\
0 & 0 & 1 & 0 & 0 & 0  
\end{pmatrix}} \ ,
\eea
which now interchanges $y^2$ with $\wy_{2}$ in the doubled coordinates $\bX$
and all fields, leaving all other components invariant. The
transformed doubled metric is given by
\bea
\cH_Q = \cO_{y^2}^\top \, \cH_f \, \cO_{y^2} = 
\begin{pmatrix}
(2\pi\,r)^2 & 0 & 0 & 0 & 0 & 0 \\
0 & \frac{A}{\tau_2(x)} & 0 & 0 & 0 & 
\frac{A\,\tau_1(x)}{\tau_2(x)} \\
0 & 0 & \frac{(2\pi\,\alpha')^2}{A\,\tau_2(x)} & 0 &
-\frac{(2\pi\,\alpha')^2\,\tau_1(x)}{A\,\tau_2(x)} & 0 \\
0 & 0 & 0 & \big(\frac{\alpha'}r\big)^2 & 0 & 0 \\
0 & 0 & -\frac{(2\pi\,\alpha')^2\,\tau_1(x)}{A\,\tau_2(x)} & 0 &
\frac{(2\pi\,\alpha')^2\,|\tau(x)|^2}{A\,\tau_2(x)} & 0 \\ 
0 & \frac{A\,\tau_1(x)}{\tau_2(x)} & 0 & 0 & 0 &
\frac{A\,|\tau(x)|^2}{\tau_2(x)}
\end{pmatrix} \ ,
\label{eq:cHQ}\eea
while the Wess--Zumino three-form is
\bea
\cK_Q = -\tfrac12\, \delta^{ac}\, M_c{}^{b}\, \dd x\wedge\dd\wy_a\wedge \dd \wy_b \ .
\eea
Reading off the closed string metric and $B$-field from
\eqref{eq:doubledmetric} yields precisely \eqref{eq:TfoldgBelliptic}
for the non-geometric $T^2$-bundle over $S^1$,
while \eqref{eq:O33dilaton}, with vanishing dilaton on the twisted
torus, yields the anticipated dilaton field
\eqref{eq:Tfolddilatonelliptic}. 
The $H$-field strength is given by $H=\dd B$. In this case the generators $\wJ^m$
generate a subgroup $\wG_\IR$ which is not preserved by $\ccG_\IZ$,
so that the quotient $\wG_\IR\setminus\ccX$ is locally modelled on the
coset $\wG_\IR\setminus\ccG_\IR$ but is not globally
well-defined, and a global description of the background in terms of conventional
geometry is not possible. However, the T-fold \emph{is} a
submanifold of the doubled twisted torus $\ccX$, because $O(2,2;\IZ)\subset
GL(4,\IZ)$ acts geometrically on the doubled torus fibres. As the only non-vanishing structure
constants are $Q_x{}^{ab}=\delta^{ac}\, M_c{}^{b}$, we obtain the allowed D2-brane
configuration displayed in
Table~\ref{tab:orbits}~\cite{Albertsson2008}; as Morita duality acts
entirely within the noncommutative Yang--Mills theory on the D2-brane,
and in particular does not mix gauge theory modes with string winding
states, the same picture of a parameterised family of D2-brane gauge
theories fibred over the $x$-circle emerges in the doubled geometry,
returning to itself under a monodromy $x\mapsto x+1$ up to Morita
equivalence, which is a symmetry of the noncommutative gauge theory. On the other hand, the
D3-brane considered in \S\ref{sec:Tfoldparabolic} is not a consistent
worldvolume when embedded as a three-dimensional subspace of the six-dimensional
doubled twisted torus $\ccX$~\cite{Albertsson2008}.

\section{D3-branes on essentially doubled spaces\label{sec:Rfolds}}

Recall that one motivation for turning to the doubled twisted torus formalism is
that it enables us to perform the  generalised T-duality transformation along the non-isometric base direction of
the original $T^2$-bundle over $S^1$. 
This maps the original D1-brane configuration to a
D3-brane wrapping an essentially doubled  
 background.
 In this final section we will discuss how to make sense of the noncommutative supersymmetric Yang--Mills theory on D3-branes in
 essentially doubled  spaces
  in the decoupling limit using the doubled
twisted torus formalism. 

\subsection{Worldvolume geometry}

To carry out T-duality along the vector field $\frac\partial{\partial x}$ of $\ccX$, we apply   \eqref{eq:O33transfa} to \eqref{eq:cHQ}
with the $O(3,3;\IZ)$ operator \eqref{eq:Ox},
which interchanges $x$ with $\wx$ in the doubled coordinates $\bX$
and all fields, leaving all other components invariant. The
transformed doubled metric is given by
\bea
\cH_R = \cO_{x}^\top \, \cH_Q \, \cO_{x} = 
\begin{pmatrix}
\big(\frac{\alpha'}r\big)^2 & 0 & 0 & 0 & 0 & 0 \\
0 & \frac{A}{\tau_2( \wx )} & 0 & 0 & 0 & 
\frac{A\,\tau_1( \wx )}{\tau_2( \wx )} \\
0 & 0 & \frac{(2\pi\,\alpha')^2}{A\,\tau_2( \wx )} & 0 &
-\frac{(2\pi\,\alpha')^2\,\tau_1( \wx )}{A\,\tau_2( \wx )} & 0 \\
0 & 0 & 0 & (2\pi\,r)^2 & 0 & 0 \\
0 & 0 & -\frac{(2\pi\,\alpha')^2\,\tau_1( \wx )}{A\,\tau_2( \wx )} & 0 &
\frac{(2\pi\,\alpha')^2\,|\tau( \wx )|^2}{A\,\tau_2( \wx )} & 0 \\ 
0 & \frac{A\,\tau_1( \wx )}{\tau_2( \wx )} & 0 & 0 & 0 &
\frac{A\,|\tau( \wx )|^2}{\tau_2( \wx )}
\end{pmatrix} \ .
\label{eq:cHR}\eea
Comparing with \eqref{eq:doubledmetric} formally gives the closed string metric
and $B$-field
\bea
g_R &=& \Big(\frac{\alpha'}r\, \dd x\Big)^2 + \frac{\tau_2( \wx )}{|\tau( \wx )|^2}\,
\Big(A \, \big(\dd y^1\big)^2+\frac{(2\pi \, \alpha')^2}{A}\,
\big(\dd y^2\big)^2 \Big) \ , \nonumber \\[4pt]
B_R &=& \frac{\tau_1( \wx )}{|\tau( \wx )|^2} \ \dd y^1\wedge \dd y^2 
\label{eq:RfluxgB}\eea
in the
essentially doubled  space,
 while \eqref{eq:O33dilaton}
yields the dilaton field
\bea
\e^{\phi_R( \wx )} = \bigg(\frac{(\alpha')^2 \, \tau_2( \wx )}{r^2 \, A\, |\tau( \wx )|^2}
\bigg)^{1/2} \ .
\eea
The explicit dependence on the dual coordinate $\wx$  reflects the
non-geometric nature of the essentially doubled  space: In this
polarisation the generators $\wJ^m$ do not close to a subalgebra  and a conventional description of the
background cannot be recovered even locally.

In this polarisation the Wess--Zumino  three-form
\bea
\cK_R = -\tfrac12\, \delta^{ac}\, M_c{}^{b}\, \dd\wx\wedge\dd\wy_a\wedge\dd \wy_b
\eea
vanishes as required on the worldvolume of the D3-brane, which wraps
the directions with coordinates $(x,y^1,y^2)$. It is shown
by~\cite{Hull2009} that it is possible to use the
self-duality constraint \eqref{eq:selfduality} to completely remove
the dependence of the doubled worldsheet sigma-model on the pullbacks
of $\dd\wx_m$ and write the doubled theory as a non-linear sigma-model
for the metric $g_R$ and $B$-field $B_R$ in \eqref{eq:RfluxgB},
depending explicitly on the winding coordinate $\wx$, thus rendering the coordinate fields non-dynamical along the dual directions $(\wx,\wy_1,\wy_2)$. 

Using \eqref{eq:openstringGTheta} we now compute the open string
metric and noncommutativity bivector on the D3-brane in the $R$-flux
background to find
\bea
G_R &=& \Big(\frac{\alpha'}r \, \dd x\Big)^2 + \frac A{\tau_2( \wx )} \, \big(\dd y^1\big)^2
+ \frac{(2\pi \,\alpha')^2}{A\, \tau_2( \wx )}\,\big(\dd y^2\big)^2 \ ,
\nonumber \\[4pt] \Theta_R &=& \tau_1( \wx ) \, 
\frac\partial{\partial y^1} \wedge\frac\partial{\partial y^2} \ .
\label{eq:openstringRflux}\eea
Thus even the open string geometry seen by the D3-brane has a 
non-geometric dependence along the transverse $\wx$-direction to its
worldvolume in $\ccX$.

\subsection{Noncommutative Yang--Mills theory}

To find a  decoupling limit with pure gauge theory on the
D3-brane worldvolume, we note that the $T^2$-fibre parts of the open
string geometry \eqref{eq:openstringRflux} coincide with those of the
T-folds, given in \eqref{eq:openstringelliptic}, upon replacing the
base $S^1$ coordinate $x$ with its dual coordinate $\wx$. Thus the scaling limit
will involve taking $\alpha'=O(\epsilon^{1/2})$,
$A=O(\epsilon^{1/2})$ and $\tau_2^\circ=O(\epsilon^{1/2})$, with
$\epsilon\to0$ and the radii \eqref{eq:scalingradii} held fixed
exactly as previously, and in addition $r=O(\epsilon^{1/2})$ with the
base radius
\bea
\bar r_x:= \frac{\alpha'}{2\pi\, r}
\eea
finite in the zero slope limit. Then the open string metric and
noncommutativity bivector are finite in this limit and can be written
as
\bea
\dd s_R^2 &=& (2\pi\, \bar r_x \, \dd x)^2 + \dd s_{\rm D2}^2\big|_{x\to\wx} 
\ , \nonumber \\[4pt]
\theta_R &=& \tau_1(\wx)\big|_{\tau_2^\circ=0}
\ \frac\partial{\partial y^1}\wedge\frac\partial{\partial y^2} \ ,
\eea
where $\dd s_{\rm D2}^2\big|_{x\to\wx}$ is the decoupled metric of a D2-brane
wrapping the $T^2$-fibre of a T-fold with all $x$-dependence
replaced by $\wx$-dependence; these quantities can be read off from
\eqref{eq:dsD3parabolic} in the case of parabolic twists and from
\eqref{eq:dsD2elliptic} (with $m\in4\,\IZ+1$ and $\vartheta=\frac\pi2$) in the case of the $\IZ_4$ elliptic
twist. Note that in this limit, the original twisted torus $X$ and
the dual T-fold
completely degenerate to a point, even though the bivector $\Theta_R$
has no components along the $S^1$ base.

From \eqref{eq:gYMDp} with $p=3$ we can also compute the Yang--Mills
coupling on the D3-brane wrapping the essentially doubled  space
 to get
\bea
g^{\phantom{\dag}}_{\rm YM}(\wx)^2 = \bigg(\frac{(2\pi\,\alpha')^2 \, g_s^2}{r^2 \, A
  \, \tau_2(\wx)}\bigg)^{1/2} \ ,
\eea
which also generally depends on the dual coordinate $\wx$. Thus in this case the
relevant parameter to be kept finite in the zero slope limit is given
by
\bea
\bar g_s^2 := \frac{(\alpha')^2 \, g_s^2}{r^2 \, A \, \tau_2^\circ} \ ,
\eea
which requires the string coupling to scale as
$g_s=O(\epsilon^{1/2})$. Then the finite Yang--Mills coupling in the
scaling limit is given by \eqref{eq:gYMparabolic} in the case of
parabolic twists, and by \eqref{eq:gYMelliptic} (for $m\in4\,\IZ+1$ and $\vartheta=\frac\pi2$) with $x\to\wx$ in
the case of the $\IZ_4$ elliptic twist. 

Since the D3-brane wraps the $T^2$ fibres over the dual $\wx$-circle
in this case, we now obtain a parameterised noncommutative worldvolume
gauge theory, with noncommutative associative star-products of fields $f,g$
given by the Kontsevich star-product
\bea
f\, \widetilde{\star}\, g = \, \mbf\cdot\, \Big[\exp\big(\, \mbox{$\frac\ii2 \, \theta(\wx)\,
(\frac\partial{\partial y^1} \otimes \frac\partial{\partial y^2} - \frac\partial{\partial y^2}
\otimes \frac\partial{\partial y^1}) $} \, \big) (f\otimes g)\Big] \ ,
\label{eq:starRflux}\eea
which is invariant up to Morita equivalence under monodromies
$\wx\mapsto \wx+1$, in the same sense as explained in \S\ref{sec:Tfoldparabolic} and \S\ref{sec:Tfoldelliptic}. This shows that a D3-brane wrapping the
essentially doubled  space
 has a sensible low-energy effective
description, which can be understood as a noncommutative gauge theory
over a compactification of the  $\wx$-direction transverse to its
worldvolume in the doubled twisted torus $\ccX$ using Morita
duality. Following the discussion of
\S\ref{sec:Morita}, in the case of the $\IZ_4$ elliptic
monodromy the noncommutative Yang--Mills action should be augmented by
replacing the noncommutative field strength tensor
$F_{\widetilde{\star}}$ with
\bea
\widetilde{\mathcal{F}} = F_{\widetilde{\star}} + \widetilde{\Phi} \ ,
\eea
where
\bea
\widetilde{\Phi} = -\frac{\wx}{\theta(\wx)} \ \dd y^1\wedge \dd y^2 \ ,
\eea
thus exibiting further non-geometric dependence of the noncommutative
gauge theory on the winding coordinate $\wx$.

{This can be interpreted as follows. Consider, for example,
a D2-brane wrapping the $(y^1,y^2)$-directions. As discussed in \S\ref{intro}, this is secretly a
D3-brane wrapping the $(\wx,y^1,y^2)$-directions in the doubled
space (see Table~\ref{tab:orbits}). In the usual (untwisted) case, there is no $\wx$-dependence of
either the closed background $(g,B,\phi)$ or the open string moduli $(G_R,\Theta_R,g^{\phantom{\dag}}_{\rm YM})$
 and one can project to the
``physical space'' with coordinates $(x,y^1,y^2)$, obtaining a
$2+1$-dimensional supersymmetric Yang--Mills theory in $(y^1,y^2,t)$-space
as the low-energy effective description of the D2-brane. On the other
hand, in the twisted case, the open string background  has a genuine $\wx$-dependence,
and so the theory cannot be interpreted in the physical space. This results
in a worldvolume theory in the full $(\wx,y^1,y^2,t)$-space. A similar
interpretation holds for other D-branes.}

\section*{Acknowledgments}

We are grateful to Dieter L\"ust, Emanuel Malek and Erik Plauschinn for helpful discussions and correspondence. This work is supported by the COST Action MP1405 QSPACE, by the EPSRC Programme
Grant EP/K034456/1,
and by the STFC
Consolidated Grants ST/L00044X/1 and ST/P000363/1.

\appendix

\section{The Buscher construction\label{app:Buscher}}

The Buscher T-duality rules are given by
\bea
\widetilde{g}_{\iota\iota} &=& \frac{(4\pi^2\,\alpha')^2}{g_{\iota\iota}}
  \ , \nonumber \\[4pt]
\widetilde{g}_{\iota \alpha} &=& -2\pi\,\alpha'\,^2\, \frac{B_{\iota
    \alpha}}{g_{\iota\iota}} \ , \nonumber\\[4pt]
\widetilde{g}_{\alpha\beta} &=& g_{\alpha\beta}-\frac1{g_{\iota\iota}}
\,\big(g_{\iota\alpha}\, g_{\iota\beta}-(2\pi\,\alpha')^2\,
B_{\iota\alpha}\, B_{\iota\beta}\big) \ , \nonumber\\[4pt]
\widetilde{B}_{\iota\alpha} &=& -\frac{g_{\iota\alpha}}{g_{\iota\iota}} \ ,
\nonumber\\[4pt]
\widetilde{B}_{\alpha\beta} &=& B_{\alpha\beta} - \frac1{g_{\iota\iota}}\,
\big(g_{\iota\alpha}\, B_{\iota\beta}-B_{\iota\alpha}\,
g_{\iota\beta}\big) \ , \nonumber\\[4pt]
\e^{\widetilde\phi} &=& \bigg(\frac{2\pi\, \alpha'}{g_{\iota\iota}}\bigg)^{1/2} \ \e^\phi \ .
\eea
Here the index $\iota$ labels the direction of the Killing vector
$\partial_\iota$ of an isometry of the initial closed string
background $(g,B,\phi)$. Note that the T-duality relations along multiple isometric
directions are formally equivalent to the open-closed string relations
\eqref{eq:openstringGTheta} with $G=\widetilde{g}$ and $\Theta=\widetilde{B}\,^{-1}$.

\end{document}